\documentclass[fleqn,usenatbib,useAMS]{mnras}

\usepackage[T1]{fontenc}
\usepackage{ae,aecompl}
\usepackage{graphicx}	
\usepackage{amsmath}	
\usepackage{amssymb}	
\usepackage{comment} 
\usepackage{lipsum} 

\usepackage{float}
\usepackage{color}
\usepackage{caption,subcaption}
\usepackage{multirow}
\usepackage{indentfirst}
\usepackage{amsmath}
\usepackage{geometry}

\usepackage{newtxtext,newtxmath}

\newcommand{\adr}[1]{{\textcolor{red}{[ #1]}}} 

\title[Formation of Globular Clusters in Cosmological Simulations]{Small but Mighty: The Formation of Globular Clusters in Cosmological Simulations \adr{SK: It's ok, but maybe we can come up with something better.}}

\title[Formation of Globular Clusters in Cosmological Simulations]{Dynamical Investigation into the Formation of Globular Clusters in Cosmological Simulations Using Lindblad Diagrams}

\title[Infant globular clusters in phase space]{Towards a phase space depiction of infant globular clusters}

\title[Infant GCs and their formation environments]{Linking the Internal Properties of Infant Globular Clusters to their Formation Environments}
\author[Phipps et al.]{
Frederika Phipps$^{1}$\thanks{E-mail: phipps@roe.ac.uk (FP)},
Sadegh Khochfar,$^{1}$
Anna Lisa Varri$^{1,2}$
and Claudio Dalla Vecchia$^{3,4}$
\\
$^{1}$Institute for Astronomy, University of Edinburgh, Royal Observatory, Edinburgh, EH9 3HJ, UK\\
$^{2}$School of Mathematics and Maxwell Institute for Mathematical Sciences, University of Edinburgh, Kings Buildings, Edinburgh, EH9 3FD, UK\\
$^{3}$Instituto de Astrof{\'i}sica de Canarias, C/V{\'i}a L{\'a}ctea s/n, E-38205 La Laguna, Tenerife, Spain\\
$^{4}$Departamento de Astrof{\'i}sica, Universidad de La Laguna, Av. del Astrof{\'i}sico Francisco S{\'a}nchez s/n, E-38206 La Laguna, Spain
}

\date{Accepted XXX. Received YYY; in original form ZZZ}

\pubyear{2022}
\hypersetup{draft}

\begin{document}
\label{firstpage}
\pagerange{\pageref{firstpage}--\pageref{lastpage}}
\maketitle

\begin{abstract}
We investigate the formation of infant globular cluster (GC) candidates in high-resolution cosmological simulations from the First Billion Years (FiBY) project. 
By analysing the evolution of the systems in the energy and angular momentum plane, we identify the redshift at which the infant GCs first became gravitationally bound, and we find evidence of radial infall of their gaseous and stellar components. The collapse appears to be driven by internal self-gravity, however, the initial trigger is sourced from the external environment. The phase space behaviour of the infant GCs also allows us to identify some characteristic groupings of objects. Such a classification based on internal properties appears to be reflected in the formation environment: GC candidates that belong to the same class are found in host galaxies of similar morphology, with the majority of the infant GCs located in clumpy, irregular proto-galaxies. 
Finally, through the inspection of two GC candidates that contain only stars by $z = 6$,  we find that supernova feedback is the main physical mechanism behind their dearth of gas and that the systems subsequently respond with an approximately adiabatic expansion. Such infant GC candidates already resemble the GCs we currently observe in the local Universe. 

\end{abstract}

\begin{keywords}
galaxies: formation -- galaxies: high-redshift -- galaxies: star clusters: general -- globular clusters: general
\end{keywords}



\section{Introduction}
\label{sec:intro}
The path through which globular clusters (GCs) are formed, despite being the subject of multiple observational and theoretical studies \citep[e.g., see][]{vanzella17,vanzella19,kruijssen15,ma2020}, currently remains unknown.
Observational evidence based on stellar populations in present-day GCs suggests their formation can take place from $z > 6$ all the way down to $z=3$ \citep[e.g., see][]{katz13,kruijssen15}.
Any plausible formation scenario has to satisfy a number of observational constraints; e.g.  
the existence of multiple populations in GCs \citep{lardo11,piotto15,milone17,bastian18}, the observed colour bi-modality \citep{larsen01,peng06,renaud17} as well as the split in the age-metallicity relation \citep{forbes10,leaman13,recioblanco18}. The latter two constraints have lead to the interpretation that
a fraction of GCs form in-situ within their host galaxy, whilst another fraction might have been accreted from in-falling dwarf galaxies \citep{cote98,tonini13,renaud17,recioblanco18}. From the colour bi-modality, GCs can be split broadly into two categories. 
The first class of GCs, 
so-called `blue', are metal-poor, do not show any net rotation as a population, and are believed to be accreted from low-mass galaxy systems.
The second class, the so-called `red' GCs, are metal-rich, and are often believed to have formed in-situ within the host that they are orbiting. 

Whilst the in-situ vs. ex-situ operational framework can not provide the fine details on the exact formation mechanism of these stellar systems, it provides empirical evidence that there could be at least two different channels of formation for GCs. If the blue GCs did indeed originate in dwarf galaxies, then, the environment they formed in could be different to those of red GCs. When considering the metallicities of the blue and red populations, it is deduced that the blue GCs formed earlier  than the red GCs, as inferred from the lack of metals in the former. 
Some early theories already attempted to address this constraint imparted by colour bi-modality: for example, \citet{schweizer87} and \citet{ashman92} presented a two-step formation channel in order to account for the age difference between the red and blue GCs. They suggested that the blue GCs formed in the early Universe, in metal-poor environments, before the formation of the first galaxies \citep[see also][]{peebles68}. These early, blue GCs then enrich the gas in their host galaxies with metals, and the galaxies themselves eventually merge with other hosts, triggering starbursts and the formation of the red GCs.


Another two-step formation channel for GCs has been proposed by \citet{forbes97}. 
The first step occurs during the collapse of the proto-galaxies, when the blue GCs are pinpointed to form. This phase likely ends when reionisation occurs, as this process quenches star formation \citep[e.g., see][]{beasley02,2013MNRAS.429L..94P}. The second step is identified as the phase when GCs formation is then resumed, once the interstellar medium (ISM) becomes dense enough due to the formation of galactic discs. This scenario naturally accounts for the observed bi-modality.  

In more recent years, it has been suggested that the GC population is just a consequence of standard clustered star formation in high redshift environments \citep[e.g., see][]{kruijssen15,keller20}, i.e. stars and star clusters form from gravitationally bound gas in giant molecular clouds (GMCs) 
\citep[e.g., see the reviews by][]{mckee07,dobbs14}. 
It has been argued that the early stages of GC formation might be similar to those of GMC formation \citep{howard18}. GMCs typically form either in a ``bottom-up" or ``top-down" process. The bottom-up process involves a series of inelastic collisions of cold clouds of HI. The mean size and mass of the overall cloud then increases gradually until it reaches the size of a GMC \citep{field65b,kwan79}. 
The main obstacle to this process are the timescales involved in reaching appropriate GMC masses. 
The ``top-down" process, on the other hand, takes place on shorter timescales and it involves large-scale instabilities in a diffuse ISM, such as the Parker instability \citep{parker66} or the Jeans instability \citep{elmegreen79,elmegreen95}.
Recent zoom-in simulations of high-redshift GC formation by \citet{ma2020} found evidence of signatures of both the bottom-up and top-down scenarios.

The formation theories summarised above are unable to explain the presence of multiple populations within GCs, which is a property present for all GCs and therefore must be intrinsic to the small-scale formation details, e.g. star-by-star formation. Whilst several mechanisms through which multiple populations can form have been presented, \citep{prantzos06,decressin07a,decressin07b,dercole08,bastian13,dercole16,bekki17,elmegreen17a}, in this study we will focus only on the fundamental dynamical aspects of the formation of infant GCs. 

From an observational perspective, whilst there is promising evidence of the existence of very compact objects at high redshift  \citep[see][among others]{vanzella17,vanzella19,vanzella20,mevstric22,elmegreen17b,bouwens17, bouwens21,bouwens22,kawamata18,kikuchihara20}, the current data sets do not yet have the required depth or detail needed to provide constraints on different formation scenarios for GCs. However, numerical simulations are an alternative way to study and assess these formation channels. Various studies have been conducted using a wide range of simulations from idealised small scale \citep[e.g.][]{nakasato00,bate03} to cosmological settings \citep[e.g.][]{kravtsov05,prieto08}. These numerical studies also come with their caveats. Many rely on local Universe constraints in order to select candidates at high redshift within their simulations \citep{halbesma20,creasey19,ma2020}. This can lead to promising candidates being identified, however, an  agnostic analysis of high-redshift cosmological simulations \citep{phipps20} can uncover GC formation channels independently of any bias stemming from their subsequent evolution.

In this work, we will focus on understanding the physical processes behind the formation of infant GCs at $z\geq 6$ by using a high-resolution cosmological simulation suite known as the First Billion Years Project (FiBY). 
In a previous article (\citealt{phipps20} - P1 from now on) we identified as `infant GC candidates' a population of stellar systems characterised by a low dark matter (DM) fraction, a high stellar density, and small half-mass radii compared to the general population of low-mass proto-galaxies present in the FiBY simulations.
The low DM content of these infant GC candidates raises questions to how the formation of these dense objects proceeded at high redshift. Since there is no underlying DM structure for the infant GCs, instabilities in the baryonic component must drive the formation. 
Whilst there are studies which are complementary in terms of spatial resolution \citep[e.g.][]{ricotti16}, they do not match the cosmic volume or the physically rich nature of the sub-grid models used in FiBY. Vice-versa, there are simulations of larger cosmic volume \citep{kim18,pfeffer18,kruijssen19,li19} but they simply cannot match the resolution achieved with FiBY. This study, therefore, provides a complementary view to those efforts.


\section{Simulations}
\label{sec:sims}

The suite of simulations used in this investigation are known as the First Billion Years (FiBY) project. These high-resolution cosmological hydrodynamic simulations were performed using a modified version of GADGET-3 \citep{spring05,schaye10}, and they include several physical processes relevant to the formation of galaxies in the high-redshift Universe, such as molecular hydrogen formation and cooling \citep[e.g., see][for more details]{johnson13}. The simulations cover the first billion years of the Universe and, therefore, end at $z=6$.  
This study uses a simulated box with a volume of $(4 $ Mpc$)^3$, that contains $2 \times 684^3$ particles. At $z = 6$, the gravitational softening length is $\epsilon = 33$ pc in physical units and 234 pc in co-moving units. The softening length is chosen to be constant in co-moving units and given in  physical units as a function of redshift as $\epsilon = 234 \times 1/(1+z)$ pc. The gas (SPH) and dark matter (DM) particles have a mass of 1250 and 6160 $\rm M_{\odot}$, respectively. This set-up operationally meets the requirements to study small-scale stellar structures whilst still appreciating the impact of the large-scale environment. The density threshold for star formation is set at $n = 10$ $\rm cm^{-3}$, and a pressure law \citep{schaye08} that yields results consistent
with the empirical Schmidt-Kennicutt Law \citep{schmidt59,kennicutt98}, is applied. The simulations take into account the formation of both Population II and III stars; the latter have an IMF modelled as a power law with a Salpeter slope \citep{salpeter55} for the mass range $21 - 500$ $\rm M_{\odot}$, while the former adopts a \citet{chabrier03} IMF. At a ``critical metallicity'' ($Z_{\rm crit} = 10^{-4}$ $\rm Z_{\odot}$), the stellar IMF transitions from that of Population III to II stars \citep{maio11}. Such a critical value is chosen on the  basis of the metallicity inferred for the most metal-poor stars to date \citep[e.g., see][]{frebel07,caffau11}. Stellar feedback is applied for both populations of stars and it is modelled via an injection of thermal energy from star particles to the neighbouring particles \citep{dallavecchia12}. 
Both Type II SNe and PISNe are included: for Population II/III stars, they occur for progenitors with an initial stellar mass of $8 \lesssim M_* \lesssim 100$ and $140 \lesssim M_* \lesssim 260$. We assume that Type II SNe and pair-instability SN inject $10^{51}$ erg and $3 \times 10^{52}$ erg, respectively. This injection of thermal energy occurs once the star particle reaches an age of 30 Myr. One SN is injected per star particle, however, the star particles are continuously releasing metals into the surrounding medium. The abundances of metals to be mixed into the surrounding medium are calculated using the tabulated yields of asymptotic giant branch stars, Type Ia and Type II SNe \citep{tornatore07,wiersma09b}. The mixing is modelled by transferring the metals to the neighbouring SPH particles in fractions weighted by the SPH kernel.
The elements tracked in the FiBY simulations are H, He, C, N, O, Ne, Mg, Si, S, Ca and Fe. Line-cooling in photoionisation for these elements uses tables from CLOUDY \citep[v07.02,][]{ferland98,wiersma09}.
Further non-equilibrium primordial chemistry networks \citep{abel97,galli98,yoshida06}, including molecular cooling for $\rm H_2$ and HD, are incorporated into the simulations. For the full details on the fundamental physical processes implemented in the FiBY project, we refer the reader to, e.g., \citet{johnson13,agarwal14,paardekooper15,cullen17}. In this investigation, we assume a  cosmology with $\rm H_0 = 71$ $\rm km s^{-1}Mpc^{-1}$, $\Omega_{\rm M} = 0.265$, $\Omega_{\Lambda} = 0.735$ and $\Omega_{\rm b} = 0.0448$.

As we are particularly concerned with the initial formation stages of the infant GC candidates as previously identified in P1, here we have focused on the analysis of a further simulation from the FiBY suite, which is characterised by a much finer time resolution than the one used in P1 (with $\Delta t \sim 2 \rm Myr$, while the main simulation used in P1 had $\Delta t \sim 50 \rm Myr$). Both simulations have the same initial conditions, box size, and mass resolution as described above. Upon a consistency check between the two simulations, we recovered the same classes of ``infant GCs'' and ``proto-UFD'' objects, as resulting from the identification prescription presented in P1 (see especially Sect.~3). The infant GC candidates from P1 that are being examined here have relatively low stellar masses compared to their local Universe counterparts, however, they are embedded in a gas rich cloud. This cloud is resolved by 1000s of particles and therefore cannot be considered a numerical artefact. 

\begin{figure*}
\includegraphics[width=\textwidth]{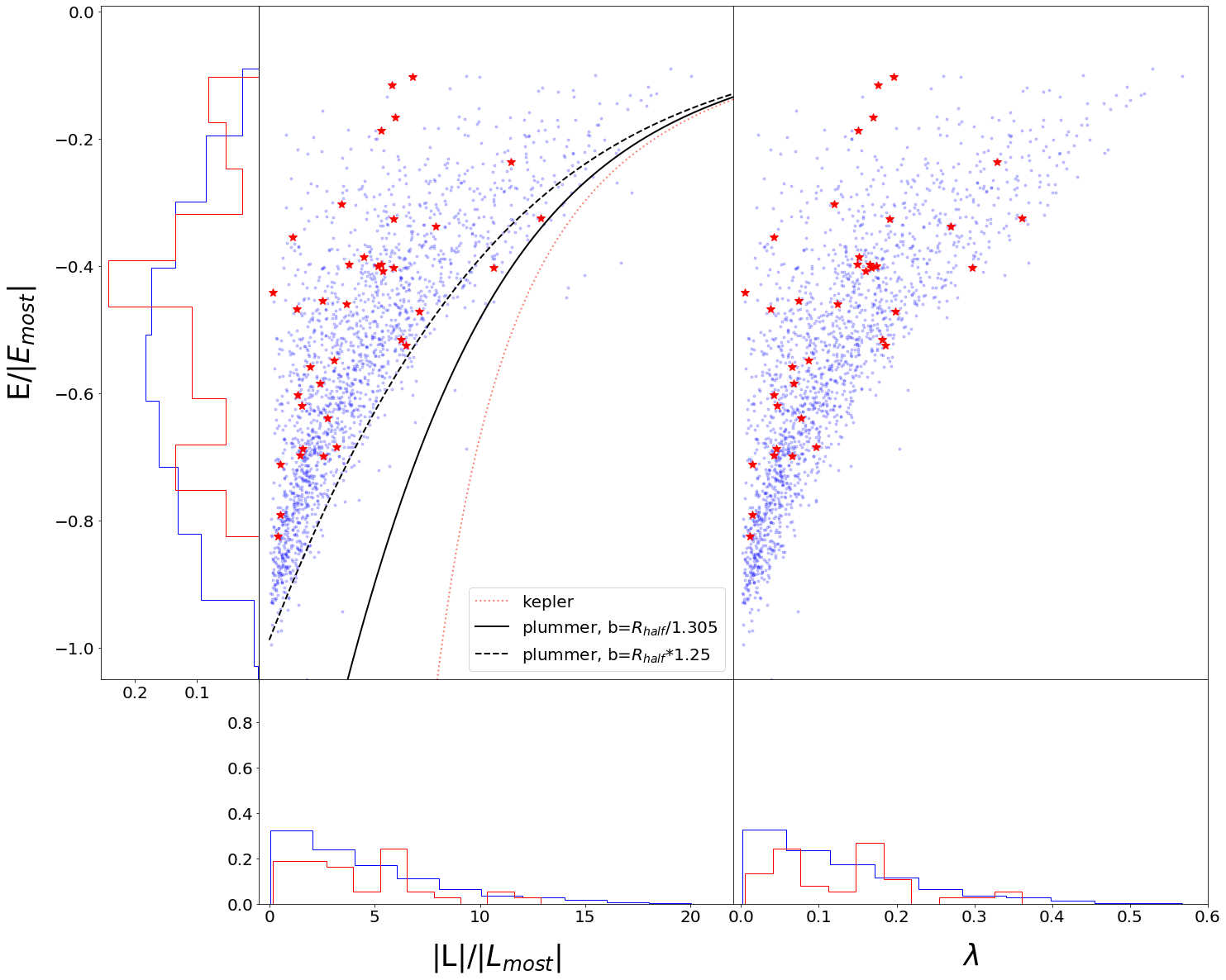}
\caption{\textit{Left:} Representation of a GC candidate in the energy-angular momentum plane (``Lindblad diagram''); for details on the adopted normalisation, see main text. Gas particles are represented by blue points whilst stars are shown with red star symbols. 
The solid, dashed and dotted lines represent the maximum allowed angular momentum for given energy in three different spherical potentials (for further details, see Section 3.1). \textit{Right:} Alternative diagram in which the angular moment is expressed by means of the dimensionless spin parameter \citep{bullock01}. On the axes, we also show histograms of the distributions of the particles by type, denoted with the same colour scheme.} 
\label{fig:redsixlindblad}
\end{figure*}

\section{Globular Cluster formation in Phase Space}
\label{sec:definitons}


Halo finders typically concern themselves with identifying matter that is gravitationally bound together in a single substructure, close to virialisation \citep[e.g.][]{spring01,knollmann09,knebe11,behroozi13}. In our previous work (P1), the substructures were identified through the SUBFIND algorithm \citep{spring01,dolag09} and followed through time using a merger tree \citep{neistein11}.  Once such structures are identified, a list of their properties are given, with particular attention to their mass and radius. While this basic characterisation is informative at the final stages of formation, it will lack a clear identification of the various components that are involved in the formation of GCs; in particular, the baryonic material that is subject to feedback and dissipational effects may be difficult to capture. 
To overcome this limitation, here we will take into account all the matter that is present in a given region, the identification of which is initially guided by the location of the particles that are bound at $z=6$.    
Equipped with this information, we will explore the formation of bound structures by studying the time evolution in the energy and angular momentum plane of all relevant matter components, as detailed below. 

\subsection{Energy-Angular Momentum Diagram}
\label{sec:lindblads}
The main tool adopted in this analysis is a representation of the phase space of a substructure of interest in terms of the energy and modulus of the angular momentum vector of its individual constituents (sometimes referred to as ``Lindblad diagram''). A characteristic example of the diagram of one infant GC candidate identified in P1 can be seen in the left-hand panels of Figure \ref{fig:redsixlindblad}, where gas and stellar particles are plotted as blue dots and red stars. The particles illustrated in this diagram are located within a sphere of radius $R_{\rm sph}= R_{\rm half}$, where $ R_{\rm half}$ is the half-mass radius of the candidate at our final redshift of $z=6$. The sphere is centered on the center of mass of the system, which, in turn, is calculated through a convergence method. We initially identify a GC candidate, its associated particles, and the relevant centre of mass by using SUBFIND. We then recalculate the center of mass by using all particles (within a sphere of radius $R_{\rm sph}$) that have negative total energy. We repeat this process iteratively, each time calculating the percentage difference between the calculated center of masses. Once this difference is 10\% or less, we consider the center of mass as converged. 
The total energy ($E$) of each particle is defined as the sum of the kinetic energy and the gravitational potential energy. The kinetic energy is given by: 
\begin{equation} \label{eqn:Ek}
E_{k} = \frac{1}{2}m_{i}v_{i,{\rm{rel}}}^2
\end{equation}
where $m_{i}$ is the mass of a given particle and $ v_{i,{\rm rel}}$ is the magnitude of the velocity vector of the particle relative to the center of mass of the infant GC candidate. We calculate the gravitational potential energy as the sum of the two-body interactions of a particle with all other particles within a given sphere of radius $ R_{\rm sph}$, centred on the center of mass
\begin{equation} \label{eqn:Ep}
E_{\rm {grav}} = -G\sum_{i,j,i\neq j}\frac{m_im_j}{r_{ij}}
\end{equation}
where $r_{ij}$ is the magnitude of the distance vector between particles i and j. As illustrated in Figure \ref{fig:redsixlindblad}, for each GC candidate we normalise the total energy of the particles to $ |E_{\rm most}|$, i.e. the average energy of the 10 most bound particles within the sphere of radius $ R_{\rm sph}$, as obtained at the end of the convergence method described above\footnote{Such a normalisation has been chosen by analysing the average energy of the most bound particles as a function of the number of particles considered; for all GC candidates, an average over $N=10$ particles allowed to identify numerically stable normalisation values.     }. 
As the second dimension of the Lindblad diagram, we consider $|L|$, i.e. the magnitude of the total angular momentum vector of the individual particles, calculated in the same frame of reference centered on the centre of mass of the GC candidate. 
For consistency with the normalisation adopted for the energy, we also normalise $|L|$ of each particle to $|L_{\rm most}|$, i.e. the magnitude of the average angular momentum of the 10 most bound particles within the sphere of radius $ R_{\rm sph}$. 

From a dynamical point of view, in the case of a system characterised by spherical symmetry, the Lindblad diagram as defined above represents a complete catalog of the orbits that are supported by the corresponding mean-field potential. All orbits must lie within the region enclosed by the vertical axis $ |L| = 0$, 
which represents the radial orbits, and the curve denoting the angular momentum and energies for a circular orbit ($L_{\rm c}$ and $ E_{\rm c}$); beyond such a ``centrifugal barrier'', which, under these assumptions, is a function of the spatial derivative of the potential under consideration, the diagram is formally inaccessible. 
For illustrative purposes, in Figure \ref{fig:redsixlindblad} we include three different options for the $L_{\rm c} - E_{\rm c}$ barrier. Two are determined by considering a \citet{plummer11} potential
\begin{equation} \label{eqn:plummer}
\Phi = - \frac{GM}{\sqrt{r^2 + b^2}},
\end{equation}
with appropriate values of the scale radius $b$.  
In the first case, which is shown as a solid black line, the scale radius is determined by the half-mass radius of the system, i.e. $b = R_{\rm half}/1.35$ \citep[for a derivation of this approximate expression, see][]{heggie03}. In the second case (dashed black line), the value of the scale radius is such that corresponding centrifugal barrier encloses the majority of the particles, as illustrated on the diagram. Finally, the third curve corresponds to a Keplerian potential, which is fully specified by the total mass $M$, and, therefore, has no further spatial degree of freedom available in the determination of the associated centrifugal barrier. We wish to emphasise that this characterisation of the $L_{\rm c} - E_{\rm c}$ curves in terms of analytical potentials is not intended to be realistic and it has been introduced exclusively to offer a simple parameterisation of the energy-angular momentum distribution of a GC candidate at the final redshift $z=6$; this will provide a convenient visual reference in the analysis of the time evolution of such diagrams, which will be presented in the next subsection.


\begin{figure*}
\centering
\includegraphics[width=0.49\textwidth]{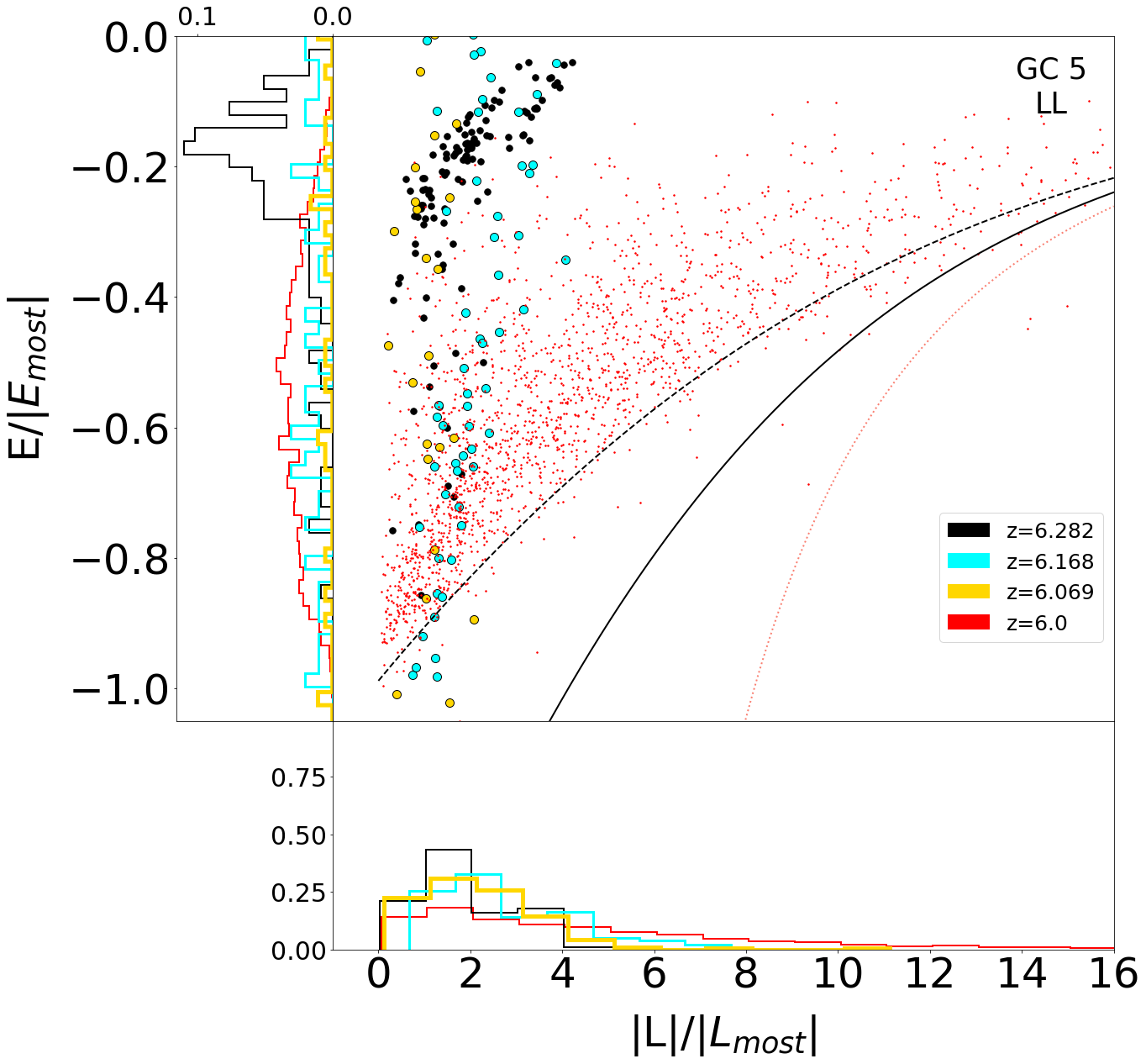}
\includegraphics[width=0.49\textwidth]{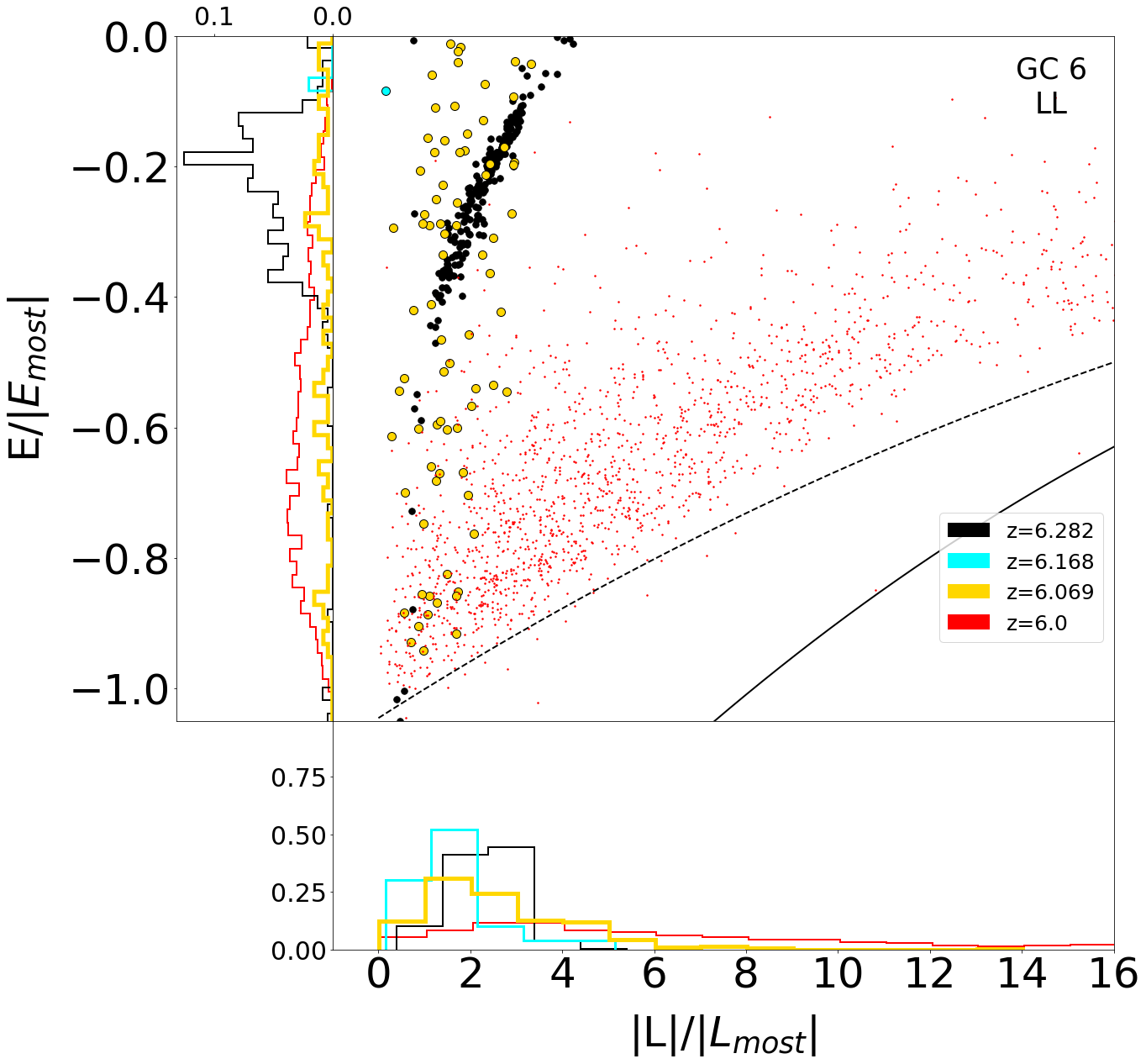}
\includegraphics[width=0.49\textwidth]{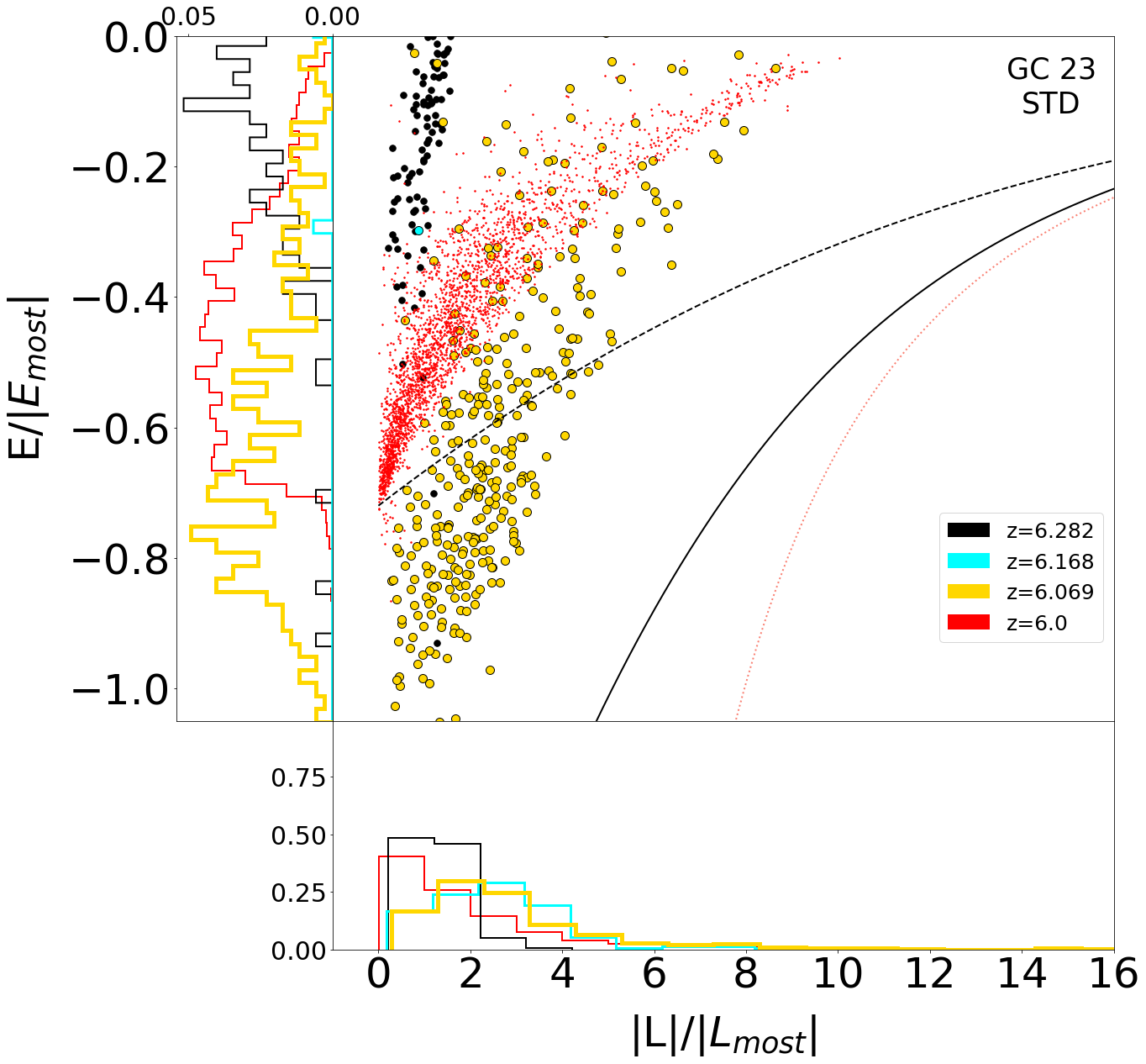}
\includegraphics[width=0.49\textwidth]{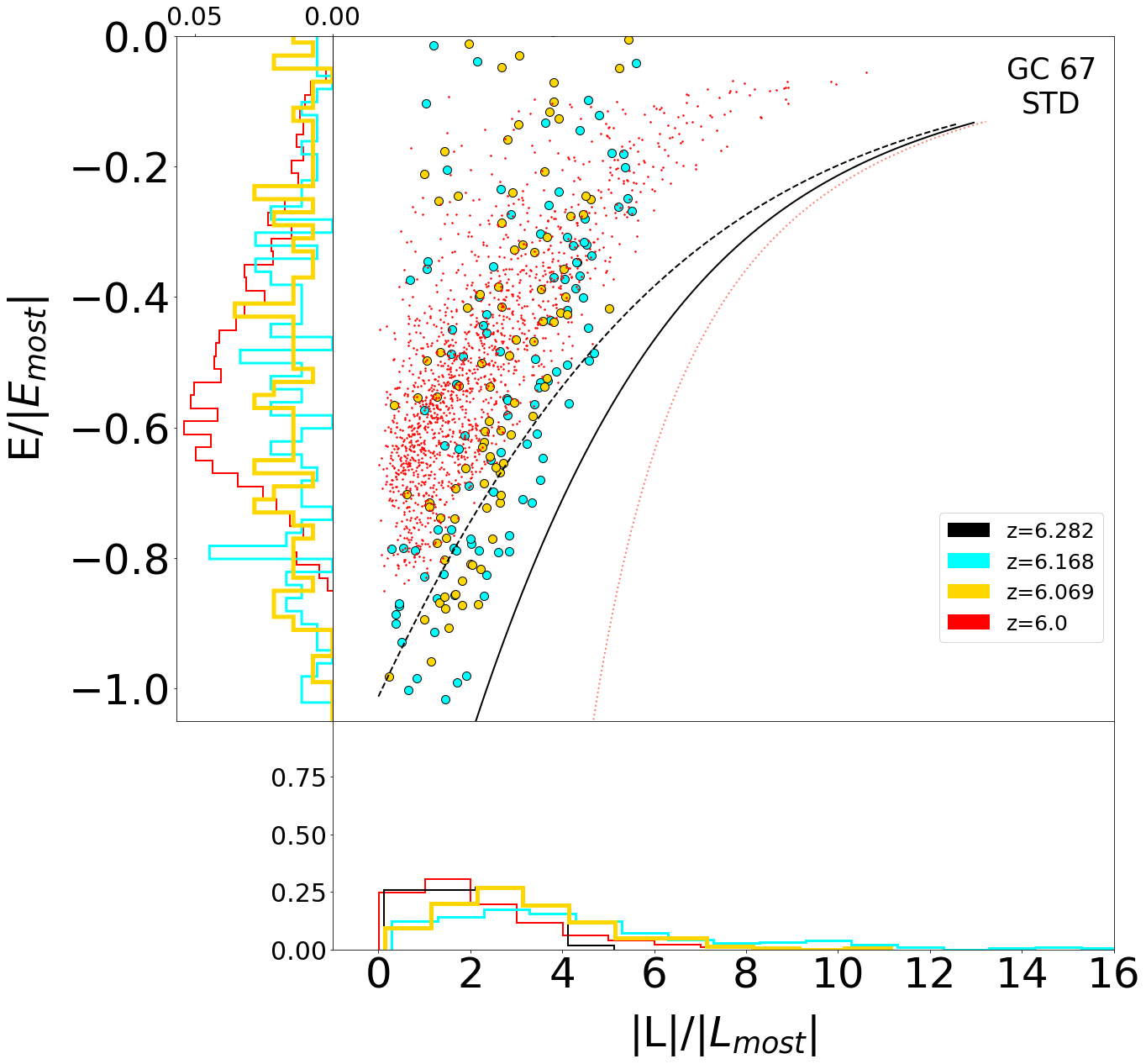}
\caption{Time evolution of the Lindblad diagram for GC candidates 5, 6, 23 and 67. Particle energy and angular momentum are normalised to the corresponding average of the 10 most bound particles at each redshift. Displayed are only the gas particles that are bound to the object at $z=6$; different colours indicate different redshifts - see the legend. From these diagrams, it appears that the gas is on nearly radial orbits suggesting infall at early times. After time has passed, the system tends to become isotropic, as the angular momentum of the gas spreads out (see x-axis histograms) and the distribution in the energy of the gas settles into the configuration we observe at $z=6$.} \label{fig:lindevol}
\end{figure*}

\begin{table*}
\centering
\begin{tabular}{cccccccccccc}
\hline
GC ID & $z_{\rm{form}}$ & $M_{T}$ & $M_{G}$ & $M_{J}$ & $R_{J}$ & $\rm frac_{DM}$ & $\rm R_{half}$ & $\log(Z/Z_{\odot})$ & $\lambda_{global}$ & Star Formation? & Group \\
& & [$\rm M_{\odot}$] & [$\rm M_{\odot}$] & [$\rm M_{\odot}$] & [pc] & & [pc] & & & & \\
\hline \hline
5 & 6.042 & $5.37 \times 10^5$ & $3.85 \times 10^5$ & $5.43 \times 10^5$ & 69 & 0.26 & 16.60 & -0.93 & 0.003 & Y & LL \\
6 & 6.042 & $1.72 \times 10^6$ & $1.24 \times 10^6$ & $5.04 \times 10^5$ & 56.9 & 0.24 & 23.30 & -1.03 & 0.003 & Y & LL \\
8 & 6.083 & $1.44 \times 10^5$ & $1.19 \times 10^5$ & $5.43 \times 10^5$ & 99.2 & 0.17 & 13.80 & -1.15 & 0.003 & N & LL \\
9 & 6.154 & $5.98 \times 10^5$ & $4.29 \times 10^5$ & $5.31 \times 10^5$ & 91.6 & 0.28 & 25.40 & -1.20 & 0.01 & Y & LL \\
15 & 6.111 & $3.57 \times 10^5$ & $3.16 \times 10^5$ & $5.26 \times 10^5$ & 76.2 & 0 & 15.80 & -1.04 & 0.002 & Y & ZDM\\
16 & 6.028 & $6.70 \times 10^4$ & $6.10 \times 10^4$ & $5.57 \times 10^5$ & 102.4 & 0 & 11.0 & -1.29 & 0.007 & Y & ZDM \\
\hline
22 & 6.125 & $3.02 \times 10^6$ & $1.66 \times 10^6$ & $4.99 \times 10^5$ & 51.3 & 0.40 & 31.90 & -1.04 & 0.001 & Y & STD \\
23 & 6.097& $6.14 \times 10^5$ & $5.20 \times 10^5$ & $5.28 \times 10^5$ & 64.2 & 0.13 & 14.40 & -1.19 & 0.0007 & Y & STD \\
24 & 6.097 & $2.12 \times 10^5$ & $1.62 \times 10^5$ & $5.99 \times 10^5$ & 132.9 & 0.23 & 19.30 & -1.22 & 0.003 & N & STD \\
25 & 6.111 & $3.24 \times 10^5$ & $3.15 \times 10^5$ & $5.62 \times 10^5$ & 104.9 & 0.019 & 23.90 & -2.16 & 0.004 & N & STD \\
34 & 6.125 & $4.73 \times 10^5$ & $3.16 \times 10^5$ & $5.50 \times 10^5$ & 99.8 & 0.33 & 19.9 & -1.24 & 0.006 & Y & STD \\
42 & 6.139 & $2.24 \times 10^6$ & $1.63 \times 10^6$ & $4.67 \times 10^5$ & 14.9 & 0.003 & 4.50 & -0.43 & 0.0001 & Y & STD \\
47 & 6.083 & $2.00 \times 10^5$ & $1.57 \times 10^5$ & $5.39 \times 10^5$ & 83.2 & 0.22 & 12.60 & -1.64 & 0.002 & N & STD \\
65 & 6.055 & $1.22 \times 10^5$ & $1.04 \times 10^5$ & $5.65 \times 10^5$ & 102.4 & 0.15 & 12.50 & -1.69 & 0.003 & Y & STD \\
67 & 6.182 & $4.01 \times 10^5$ & $3.64 \times 10^5$ & $5.99 \times 10^5$ & 142.9 & 0.09 & 31.20 &-2.46 & 0.002 & N & STD \\
\hline
10 & 6.014 & $7.84 \times 10^4$ & 0 & - & - & 0.16 & 7.90 & - & 0.013 & Y & SO \\
35 & 6.014 & $1.99 \times 10^5$ & 0 & - & -  & 0.12 & 11.60 & - & 0.033 & Y & SO \\ 
\hline
3 & >6.282 & $6.94 \times 10^7$ & $3.78 \times 10^7$ & $4.47 \times 10^5$ & 9.4 & 0.05 & 12.0 & -0.12 & 0.0004 & Y & NSC \\ 
\end{tabular}
\caption{Selected physical properties of the infant GC candidates at their `formation redshift'. They are divided into groups of objects that show similar evolution in their Lindblad diagram. From left to right: the ID of the candidate, formation redshift, total mass within $R_{\rm{half}}$, total gas mass within $R_{\rm half}$, Jeans mass, Jeans length, dark matter fraction, half-mass radius, average gas metallicity, global spin parameter (\citealt{bullock01}), whether any stars present within $R_{\rm half}$ at $z=6$ have already formed at $z_{\rm{form}}$. The final column indicates the group designation based primarily on their evolution in phase space. The groups are: nuclear star cluster ({\it NSC}), largest range in angular momentum ({\it LL}), zero-dark matter ({\it ZDM}), standard ({\it STD}) and  star-only ({\it SO}). For clarity, we find the following candidates in the same host galaxy: 5 to 10,  22 to 25, 15 \& 16 and 34 \& 35. We would like to note that the half-mass radius presented for the objects in this Table and Table \ref{table:hostproperties} for the infant GC candidates should be seen as upper limits due to the scale of the softening (for context, the softening length at $z = 6.282$ is 32 pc). The sizes here are smaller than the ones presented in P1 \citep{phipps20} as in that work we chose to utilise the radius calculated by the SUBFIND algorithm \citep{spring01} which is based on estimates of the density field. However, in this work, we decided to use the iterative method as described in Section \ref{sec:definitons} as we are interested in the most conservative definition of size to be able to identify material close to the core which allows for a more conservative tracing of the objects evolution through time to earlier redshifts. }  \label{table:formationproperties}
\end{table*}

For completeness, in the right-hand panels of Figure \ref{fig:redsixlindblad}, we display also an alternative version of the Lindblad diagram, by evaluating the total energy of particles (as defined above) against the individual spin parameter, $\lambda$. Here we have adopted the definition proposed by \cite{bullock01}, but we note that this dimensionless parameter was first introduced by \cite{peebles69} to characterise the angular momentum of bound stellar structure of cosmological relevance. 

An interesting feature of Figure \ref{fig:redsixlindblad} is the distribution of the various species of particles: the gas and star particles appear to have a similar spread in both energy and angular momentum. In this representative infant GC candidate, as in most of the others identified in this study, the number of star particles is lower compared to gas particles; this seems to suggest that gas is the dominant contribution to the local gravitational potential and that stars adjust to it. As implied by the adopted GC candidate selection criterion (see P1, Section 3), only very few dark matter particles are present in these systems. We also note that such dark matter particles, whilst having negative total energies, are usually located on the far right-hand side of the Lindblad diagram, which signifies a high angular momentum content. 
This location in the diagram suggests that they might have been initially associated to the extended dark matter halo of the host galaxy and subsequently captured gravitationally by the GC candidate. 


\subsection{Formation and evolution before $z=6$ }
\label{sec:lindbladevolution}
We can now explore the formation  and early evolution of the infant GC candidates by assessing the time evolution of their Lindblad diagram before $z=6$. 
We follow this evolution through 21 snapshots of the FiBY simulation, covering a time span of 56 Myr, which, for all our infant GC candidates, corresponds to several dynamical times. 

\begin{figure}
\centering
\includegraphics[width=0.5\textwidth]{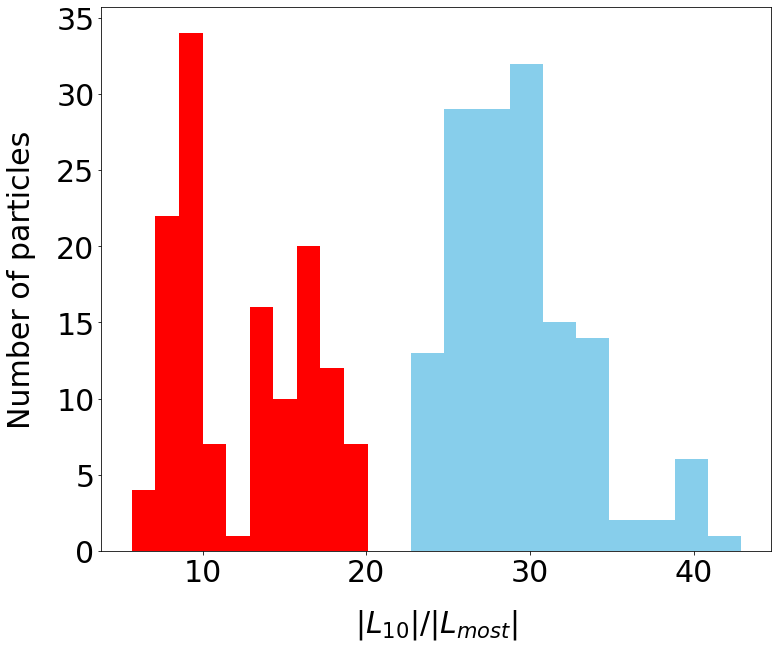}
\caption{Histogram of the top ten percent of gas particles GC per candidate with the highest angular momentum, normalised to the angular momentum of the most bound particles at $z=6$. The blue histogram represents particles in the {\it LL } group of GC candidates, whilst the red is the {\it STD} group. } \label{fig:anghistogram}
\end{figure}

To compute the Lindblad diagram through time, as displayed in Figure 2, we apply the following approach. First, for a given GC candidate, we consider the particles that, at $z=6$, are located within a sphere of radius $ R_{\rm sph}$ and we trace them back through time. At each snapshot, we use the positions of these particles at an earlier time to determine an initial center of mass as well as a half-mass radius ($\rm R_{{\rm half},z}$) for the ancestors of the objects. 
To determine the appropriate center of mass of the systems at the earlier redshifts, we apply a similar convergence method as for the analysis conducted at $z=6$; in the first step of the iteration, we choose all particles that are bound within the sphere of radius defined by the particles bound at $z=6$. After this step, we then calculate the energy and angular momentum for each particle within $R_{{\rm sph},z}$. For this work, we choose $R_{{\rm sph},z} = R_{{\rm half},z}$. Ultimately, this conservative choice is made because, at $z=6$, the particles within the half-mass radius represent the core of the infant GC candidate and can be safely associated with it. Also, this collection of inner particles will most likely evolve to become the core of the GC at later times and, therefore, it is less likely to be disrupted by external environmental effects. 

When calculating the total energy and angular momentum, we use every bound particle within a sphere of radius $\rm R_{half,z}$, however, in Figure \ref{fig:lindevol},  we display only particles that are part of the bound object at $z=6$. This visualisation choice is made so we can concentrate on examining the phase space assembly and evolution of the core itself, as it appears at the final redshift. 
As before, the axes of the diagrams presented in Figure \ref{fig:lindevol} are normalised to the 10 most-bound displayed particles at each given redshift, and the curves denote the ``centrifugal barriers'' corresponding to the spherical potentials assessed exclusively at $z=6$, as introduced in the previous subsection. For visual clarity, we show only the gas particles, as they dominate the potential of the system; at all times, the stars follow approximately the distribution of the gas particles, as demonstrated in the case of the Lindblad diagram at $z=6$ (see Figure \ref{fig:redsixlindblad}).

When examining the Lindblad diagrams as a function of time, we can identify different phases of dynamical evolution, relating to the infall and collapse of the systems.
For most instances, $\sim$ 56 Myr ago, the infant GC candidate had not yet formed but the gas is in the process of collapsing. This can be seen through the column of gas particles (black and blue data points in Figure \ref{fig:lindevol}) at very low angular momentum, but high energies. The gas particles at these early times are on nearly radial orbits, which indicate infall. Also, the little variation we see across time in angular momentum (compared to the energy of the gas) may indicate that angular momentum is conserved during the early formation stages and suffers little re-distribution.  After a few Myr, the distribution in energy still occupies a wide range (blue and yellow data points) and the distribution in angular momentum also progressively spreads out. 

We then see that, as the collapse proceeds, the distribution in energy approaches more negative values as the gas becomes gravitationally self-bound and settles into the characteristic configuration we observe at $z=6$. This behaviour in the Lindblad diagram coincides with the point at which nearly all of the gas reaches high densities, beyond the star formation threshold. The general evolution depicted in the diagram suggests that, after an initial period of radial infall, the systems tend to become more isotropic, as the distribution of the angular momentum of the gas and star particles becomes more extended. This happens on timescales well within the dynamical time of the candidates and supports the notion that the initial orbital structure of GCs is set during a short time window that coincides with the onset of star formation. We also note that there are a few particles (in all our GC candidates) that temporarily experience a large excursion in energy. The cause for this in our simulations is supernovae in the nearby environment imparting thermal energy to neighbouring SPH particles and increasing the gas particles' kinetic energy.


\begin{table*}
\centering
\begin{tabular}{ccccccccc}
\hline
GC ID & $M_T$ & $M_G$ & $\rm frac_{DM}$ & \rm $R_{\rm{half}}$ & Galaxy Morphology & Galaxy ID & Host Mass & Group \\
 & [\rm $M_{\odot}$] & [$\rm M_{\odot}$] & & [pc] & & & [$\rm M_{\odot}$] & \\
\hline \hline
5 & $2.39 \times 10^6$ & $2.34 \times 10^6$ &  0.0026 & 8.80 & Proto-Merging Spiral-like & 1 & 3.26 $\times$ $\rm 10^9$ & LL \\
6 & $2.37 \times 10^6$ & $2.32 \times 10^6$ & 0.0026 & 7.50 & Proto-Merging Spiral-like & 1 &  3.26 $\times$ $\rm 10^9$ & LL \\
8 & $1.25 \times 10^6$ & $1.22 \times 10^6$ & 0.0049 & 15.50 & Proto-Merging Spiral-like & 1 &  3.26 $\times$ $\rm 10^9$ & LL \\
9 & $8.22 \times 10^5$ & $8.10 \times 10^5$ & 0 & 10.30 & Proto-Merging Spiral-like & 1 &  3.26 $\times$ $\rm 10^9$ & LL \\
15 & $1.56 \times 10^6$ & $1.54 \times 10^6$ & 0 & 9.0 & Rotating Spiral-like & 71 & 3.19 $\times$ $\rm 10^9$ & ZDM \\
16 & $7.15 \times 10^5$ & $7.04 \times 10^5$ & 0 & 10.30 & Rotating Spiral-like & 71 & 3.19 $\times$  $\rm 10^9$ & ZDM\\
\hline
22 & $5.10 \times 10^6$ & $4.89 \times 10^6$ & 0.003 & 7.90 & Clumpy Proto-Galaxy & 235 & 2.29 $\times$ $\rm 10^9$ & STD\\
23 & $3.56 \times 10^6$ & $3.47 \times 10^6$ & 0.005 & 8.70 & Clumpy Proto-Galaxy & 235 & 2.29 $\times$ $\rm 10^9$ & STD\\
24 & $1.14 \times 10^6$ & $1.12 \times 10^6$ & 0.066 & 10.50 & Clumpy Proto-Galaxy & 235 & 2.29 $\times$ $\rm 10^9$ & STD\\
25 & $8.38 \times 10^5$ & $8.27 \times 10^5$ & 0 & 14.90 & Clumpy Proto-Galaxy & 235 & 2.29 $\times$ $\rm 10^9$ & STD \\
34 & $1.20 \times 10^6$ & $1.18 \times 10^6$ & 0.24 & 10.70 & Clumpy Proto-Galaxy & 360 & 2.14 $\times$ $\rm 10^9$ & STD \\
42 & $2.65 \times 10^6$ & $0$ & 0.04 & 6.60 & Clumpy Proto-Galaxy & 458 & 1.84 $\times$ $\rm 10^9$ & STD \\
47 & $1.38 \times 10^6$ & $1.31 \times 10^6$ & 0 & 10.80 & Clumpy Proto-Galaxy & 472 & 1.54 $\times$  $\rm 10^9$ & STD \\
65 & $7.95 \times 10^5$ & $7.21 \times 10^5$ & 0.23 & 12.70 & Clumpy Proto-Galaxy & 649 & 4.68 $\times$ $\rm 10^8$ & STD \\
67 & $2.10 \times 10^6$ & $2.04 \times 10^6$ & 0.11 & 16.40 & Clumpy Proto-Galaxy & 676 & 8.12 $\times$ $\rm 10^8$ & STD \\
\hline
10 & $4.41 \times 10^5$ & 0 & 0.15 & 7.90 & Proto-Merging Spiral-like & 1 &  3.26 $\times$ $\rm 10^9$ & SO \\
35 & $3.91 \times 10^5$ & 0 & 0.12 & 11.60 & Clumpy Proto-Galaxy & 360 & 2.14 $\times$ $\rm 10^9$ & SO \\
\hline
3 & $8.44 \times 10^6$ & $5.14 \times 10^6$ & 0.009 & 3.60 & Proto-Merging System & 0 & 3.85 $\times$ $\rm 10^9$ & NSC  \\
\end{tabular}
\caption{Selected properties of the infant GC candidates at $z=6$. They are divided into groups of objects that show similar evolution in their Lindblad diagrams. From left to right the columns show: the ID of the candidate, dark matter fraction and half-mass radius, morphology of the host galaxy, total mass of the host galaxy, and the group designation from the phase space evolution.} \label{table:hostproperties}
\end{table*}


We find that different candidates can be grouped on the basis of their evolution in the Lindblad diagram. In general, objects with a similar phase space evolution are found to belong to the same galaxy, or to be located in host galaxies of similar morphology. In Table \ref{table:formationproperties} we indicate these groupings, as identified on the basis on their Lindblad diagram evolution. 
The panels in Figure \ref{fig:lindevol} illustrate a representative candidate of each of the classes presented in Table \ref{table:formationproperties}. The last column of the Table gives the group designations, which we will expand upon here. 
When comparing the angular momentum distribution of GC candidates we find a clear separation into two distinct populations (see Figure \ref{fig:anghistogram}), one with a distribution that shows a large range in angular momentum, which we label {\it LL}. 
The other population that we find, which is more common (labelled {\it STD}), shows a smaller spread in angular momentum, and, in addition, appears to have a very distinct evolution in energy. During the collapse phase, the gas becomes more bound than the expectation based on the reference Plummer profile identified at $z=6$.  This indicates that the gas cloud ancestor of these candidates collapsed further, which is a reflection of the lower initial angular momentum of the gas cloud compared to the {\it LL } class.  
We also identify another sub-group, zero dark matter candidates (labelled {\it ZDM}), that have a similar evolution to the large angular momentum  class {\it LL}, but in contrast, no dark matter associated with their phase space evolution at any time. Again, in these cases, the initial angular momentum of the gas cloud prohibits deep collapse which is seen in the {\it STD} class.  


For completeness, we wish to mention three GC candidates that stand out from the overall sample. The first is candidate 3 (labelled {\it NSC}). This candidate is the most massive we have identified, with a total mass of $M_{T}=6.94 \times 10^7$ M$_{\odot}$ and, across the 21 snapshots we have studied, its Lindblad diagram had already settled into an evolved configuration in the energy - angular momentum plane, indicating that its formation time is much earlier than $z=6.28$. Given such phase space appearence and its mass, we speculate that this candidate might be a potential ancestor of a nuclear star cluster. 
The other two interesting candidates are 10 and 35 ({\it SO} candidates). These two candidates are the only ones that consist only of stars at $z=6$, and they appear to be gas-poor throughout most of their evolution. 
We will come back later to the {\it SO} candidates in Section \ref{sec:staronly} to discuss how feedback might have shaped their properties.
The Lindblad diagram also gives an approximate indication of the formation time of the infant globular cluster candidates observed at $z=6$. Here ``formation'' refers to the moment at which the Lindblad diagram of a given GC candidate settles into the characteristic configuration in the energy - angular momentum plane, as observed at $z=6$. It is interesting to note that this moment coincides closely with the time at which the most bound particles stay roughly the same up to $z=6$, and the majority of stars, which later are observed at $z=6$, start to form.
In Table \ref{table:formationproperties}, we list all these formation redshifts, denoted as $z_{\rm{form}}$. 

\section{Collapse of infant GC\lowercase{s}}
\label{sec:collapse}

\subsection{Collapse of the Gas}
Having established the approximate formation redshift of all infant GC candidates, we can look into how the gas in the objects collapses and characterise its overall evolution, pre- and post-formation. 
At formation, in all GC candidates, the majority of the gas particles are above the star formation density threshold adopted in the FiBY simulations and, thus, they are in the process of being converted into stars.  
This provides further evidence that $z_{\rm{form}}$ is a good proxy for the formation of a gravitationally bound infant GC, as almost all gas has reached density values which are appropriate for giant molecular clouds.

\begin{figure}
\centering
\includegraphics[width=\columnwidth]{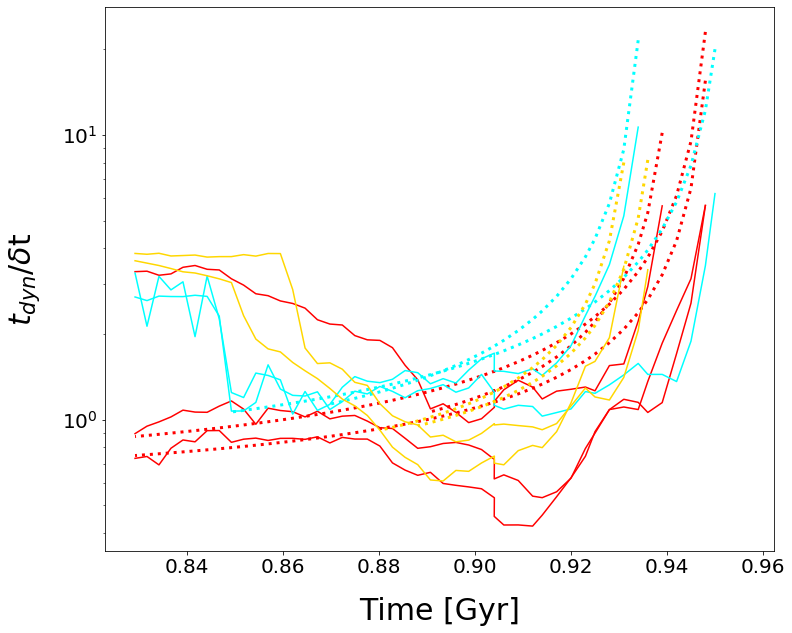}
\caption{Evolution of the ratio of the dynamical time over $\delta t$ - the time until `formation' for a selection of candidates that represent each of the main groupings: {\it LL} (red), {\it ZDM} (blue) and {\it STD} (orange). The dotted lines indicates the analytic prediction (see text). Proximity to the dotted lines indicates free-fall collapse. However, at early times some of the GC candidates show ratios greater than the analytical prediction, which suggests that other forces besides self-gravity are at play at the collapse inception.    
} \label{fig:tdyndeltat}
\end{figure}

The evolution of the gas at density values which are beyond the adopted star formation threshold is governed by an effective equation of state, to ensure the Jeans mass is always resolved. 
Therefore, to study the evolution of the gas post-formation redshift will not help us gain much insight into its collapse. However, pre-formation most of the gas has densities less than the star formation threshold, thus we can study its collapse and the evolution of its density during this time frame to gain some physical insight. 
We wish to reiterate, as already stated in the previous Section, that ``formation redshift'' simply refers to the operational definition we are adopting. In reality, the process of formation for these objects will have started at earlier times, with the initial infall of the gas; here ``formation redshift'' just marks the time at which the GC candidates can be considered as gravitationally bound objects.  
\begin{figure}
\centering
\vspace{0.8cm}
\includegraphics[width=\hsize]{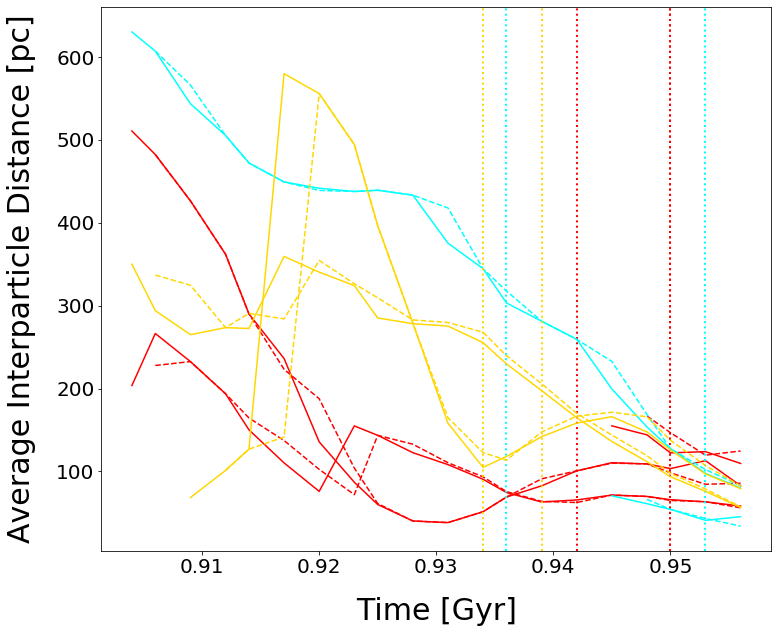}
\caption{Evolution of the average interparticle distance between the stars present in the infant GC candidates at $z=6$. All stars are shown with solid lines whilst stars that were present in the previous snapshot are shown with dashed lines. Vertical lines indicate the formation times of each of the candidates. The similarity between the solid and dashed lines shows that the new stars form consecutively closer together with time. The red ({\it LL}), blue ({\it ZDM}) and orange ({\it STD}) lines show candidates from each of the main groupings as in Figure \ref{fig:tdyndeltat}.}
\label{fig:interparticle}
\end{figure}

\begin{figure}
\centering
\includegraphics[width=0.94\hsize]{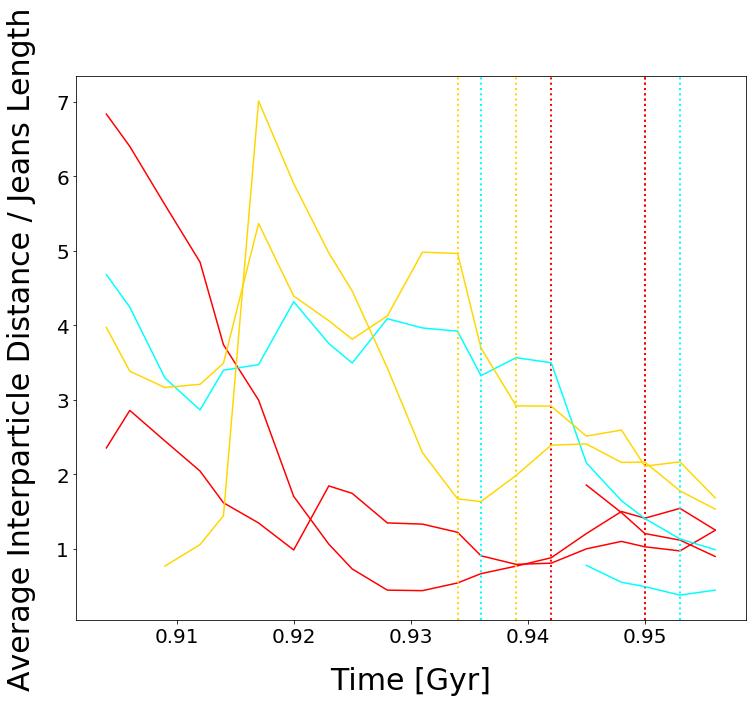}
\caption{Evolution of the ratio of the average interparticle distance between the stars present in the infant GC candidates at $z=6$ and the Jeans length of the gas. Line colours are the same as in Figure \ref{fig:tdyndeltat}. 
Vertical lines indicate the formation times of each of the candidates. The evolution of this ratio is similar to that in Figure \ref{fig:interparticle} as the Jeans length evolution is roughly constant. By $z=6$ this ratio is equal to 1, implying that the physical size of the candidates is set at early times by the Jeans length of the gas.}
\label{fig:distandjeans}
\end{figure}

The simplest way for the gas to collapse is under the influence of its own gravity. In an idealised setting where the only force at play is gravity, the gas will collapse on a free-fall timescale. Comparing a free-fall model to the dynamics of the gas pre-formation $(z > z_{\rm{form}})$ can offer us some insight into whether forces other than gravity are at work during the gas collapse. 
We compare the evolution of the gas density in the simulation to a basic free-fall model  \citep{1969MNRAS.145..271L,1969MNRAS.144..425P,omukai00} given by 
\begin{equation} \label{eqn:rateofchange}
\frac{d\langle \rho \rangle }{dt} \approx \frac{\langle \rho \rangle }{t_{\rm dyn}}
\end{equation}
where we approximate the dynamical time as $t_{\rm dyn} = 1/\sqrt{G \langle \rho \rangle }$, with G as the gravitational constant and $\langle \rho \rangle $ an average density estimate based on the density distribution of the gas in each GC  candidate. 
Specifically, we wish to confront the dynamical timescale with $\delta t$, which is the time required to reach $z_{\rm{form}}$.
As we compare these two timescales over a range of different redshifts, during periods when the collapse is purely driven by gravitational free-fall, we expect to find $t_{\rm dyn} \sim \delta t$, as dictated by the solution of Equation \ref{eqn:rateofchange}. 
Figure \ref{fig:tdyndeltat} shows as solid lines the evolution we find for the actual collapse of the GC candidates in the FiBY simulation.
In most cases, the ratio $t_{\rm dyn}$ over $\delta t$ is close to unity, which suggests that the dominant force acting on the objects is gravity and, whilst there could be some internal pressure support, it is insufficient to prevent collapse on a free-fall time. 
Also, all GC candidates 
do collapse on a timescale that is slower than $t_{\rm dyn}$, which suggests that this phase of infall is driven by the self-gravity of the system.
At the earliest times, for a 
a few GC candidates, the ratio is 
greater than the analytical prediction. This implies that the initial trigger for the collapse in these objects is not self-gravity alone, but an external process or instability sourced by the environment in the host galaxy. In consideration of the limitations of our analysis discussed in Section 7, we do not see it as appropriate to identify any particular effect as the dominant one.  We wish to note, however, that self-gravity does quickly take over and subsequently drives the collapse dynamics. 

Interestingly, the GC candidates that have a similar $t_{{\rm dyn}} / \delta t$ evolution ({\it LL} \& {\it ZDM} vs {\it STD}) are the ones which we identified earlier as being grouped on the basis of the similarity of their Lindblad diagram evolution (see Table \ref{table:formationproperties}); these groupings are displayed in Figure 4 with lines of the same colour. Although, in general, all GC candidates do undergo dynamical collapse, the subtle nuances in the time evolution of the ratio of the dynamical time over $\delta t$ are still hinting at possible differences during the collapse, for which we also see some evidence in the evolution  of the Lindblad diagrams. In particular, the low angular momentum {\it STD} population shows an evolution that continuously declines in $t_{{\rm dyn}} / \delta t$, before going through a phase of roughly constant $t_{{\rm dyn}} / \delta t$. In contrast the, {\it LL} and {\it ZDM} classes show an extended period of constant $t_{{\rm dyn}} / \delta t$.    

To further confirm that the dynamics of the collapse is consistent with the effect of self-gravity, we calculate the analytic solution of Equation \ref{eqn:rateofchange} for each candidate, by starting at a time when $t_{{\rm dyn}} / \delta t \sim 1$ and by using the average density of the GC candidate at that time. For the first tens of Myrs, the analytic solutions (displayed as dashed lines in Figure 4) are in good agreement with the numerical results from the FiBY simulation, then there is a deviation. However, such a deviation from the expectations of free-fall model is not too substantial, especially when the density of the gas at that state is taken into account. 

Even for the GC candidates that display signs of star formation at $z > z_{\rm{form}}$, this activity does not seem to affect the collapse of the gas. This lack of impact on the gas dynamics is likely due to the fact that such star formation activity 
is recent.

\subsection{Infall of Stars}

For a majority of the infant GC candidates, a fraction of the stars that, by $z=6$, are located in the core of the object has already formed by the time we consider our candidates to be gravitationally bound objects. From an examination of the time evolution of the stellar mass enclosed within a sphere of radius $R_{\rm half}$, we find that not all of the stars that exist before redshift $z_{\rm{form}}$ are formed locally within $R_{\rm half}$. 
This may be  partly due to the adopted method of determining the center of mass, as, between snapshots, the position of the barycentre fluctuates slightly. However, this effect does not account for the total infall of the stars that we observe in Figure \ref{fig:interparticle}. In this Figure, we display the interparticle distance for the star particles which are present in the GC candidates at $z=6$. When considering the stellar component of the infant GC candidates, we only show the average interparticle distance as a proxy for the extent of the stars in the object as the stellar component is resolved by a lower number of particles compared to the gas. Solid lines represent all stars, whereas the dashed lines correspond to the stars that are already present in the previous snapshot. This illustration highlights the impact on the interparticle distance of stars which are formed between snapshots. Most of the GC candidates show a constant decrease in interparticle distance over time, which can be interpreted as the particles moving toward each other during collapse. 

When comparing the time evolution of interparticle distance with the Jeans length of the gas in the GC candidates (see Figure \ref{fig:distandjeans}), we find that, initially, the Jeans length is smaller than the interparticle distance between star particles and that the ratio of interparticle distance to the Jeans length becomes roughly constant around  $z_{\rm{form}}$. The Jeans length in this Figure is calculated at each snapshot by using the average density and temperature of the gas particles within a sphere of radius $R_{\rm half}$ at that time. 
These results, partnered with evidence of the gas collapsing on the dynamical timescale, suggest that the formation of the infant GC candidates is primarily driven by gravitational instabilities. The physical size of the GC candidates in the FiBY simulation is, therefore, likely set at early times by the Jeans length and then, subsequently, by further infalling material. We emphasise that the Jeans length, as well as the sizes we calculate for our objects, should be interpreted only as upper limits due to the resolution limits of the FiBY simulation. 

In summary, through an assessment of the gas distribution over time, the star formation histories, and interparticle distance between star particles, we have determined that self-gravity dictates the timescales of the collapse of the GC candidates as well as the reference length scales they form on through the Jeans instability. The onset of gravitational instabilities in some of our cases, however, requires an external trigger that quickly becomes sub-dominant with respect to self-gravity. The limitations of our analysis (see Section 7) do not allow us to identify unambiguously the nature of such trigger, which may range from the effect of small-scale turbulence to global external tides.


\begin{figure*}
\centering
\includegraphics[width=0.9\columnwidth]{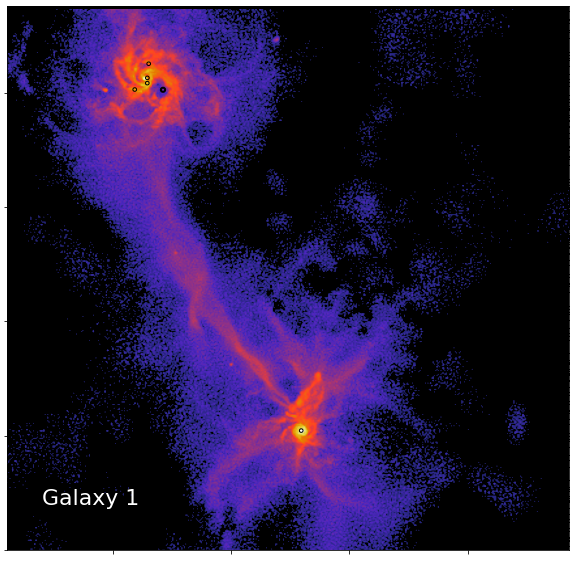}
\includegraphics[width=1.1\columnwidth]{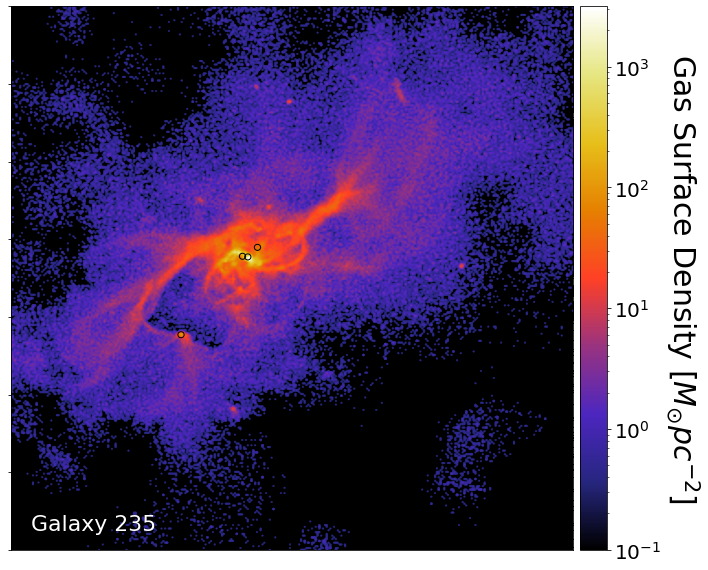}
\includegraphics[width=0.9\columnwidth]{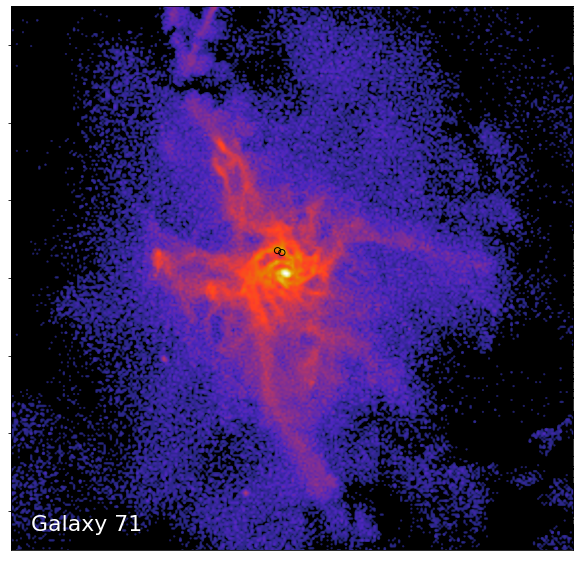}
\includegraphics[width=1.1\columnwidth]{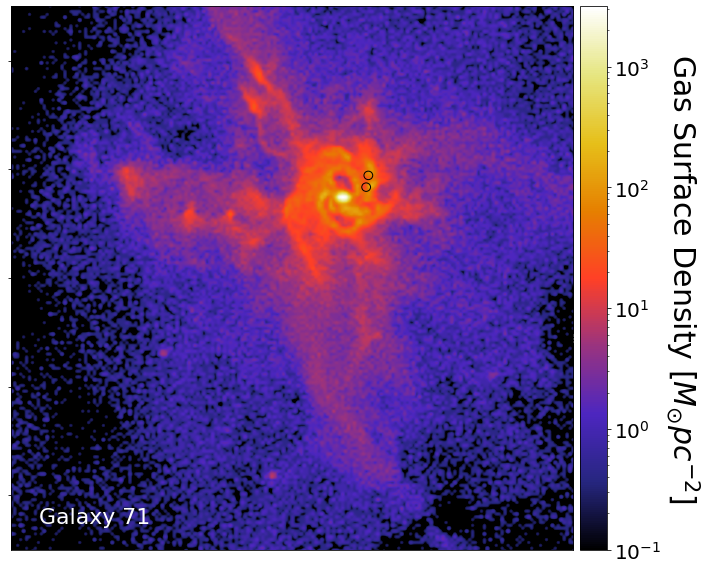}
\caption{Gas surface density maps of selected host galaxies, evaluated at the formation time of one of the GC candidates in the system. \textit{Top left:} Galaxy 1 hosts five GCs (it is merging with Galaxy 0, which hosts one candidate). \textit{Top right:} Galaxy 235 hosts four candidates. \textit{Bottom:} Galaxy 71 hosts two candidates. This system is displayed at two different redshifts, that correspond to the formation times of the two GC candidates in the galaxy.} \label{fig:hostsystems}
\end{figure*}

\section{The infant GC\lowercase{s} - host connection}
\label{sec:environment}
In the local Universe, GCs are most often found as  populations of clusters in the haloes of galaxies. As presented in P1 and confirmed in the high-frequency simulation used here, the infant GC candidates that we identify are typically located within the half-mass radius of their host galaxy. Given that most of such objects formed just recently, they are indeed still located close to their birth site. Interactions and mergers with other galaxies at lower redshift will likely relocate the surviving GC candidates to the haloes of new host galaxies. The prime epoch for these events to happen is expected to be toward $z\sim 2$, when the merger rate of galaxies peaks \citep{2001ApJ...561..517K}.    

We find that the GC candidates that belong to the same class, as attributed on the basis of their phase space properties  (see Table \ref{table:formationproperties}), also reside in hosts of similar morphology, e.g. the {\it STD} candidates are all found in clumpy host galaxies. In Figure \ref{fig:hostsystems} we present gas surface density maps for three of the GC-host galaxy systems. We refer to these systems by their parent galaxy ID: 1, 71 and 235. The location of the infant GC candidates in each of the galaxies is indicated by a black circle. 
One of the host galaxies is undergoing a major merger, whilst the infant GC candidates are forming and evolving (top left panel - Galaxy 1). In the bottom panels, we present a host (Galaxy 71) which contains two GC candidates, displayed at their corresponding formation redshifts. This host galaxy is a rotating spiral-galaxy-like system, as it can be seen by the change in orientation of the spiral arms from one snapshot to the other; the infant GC candidates are located in the spiral arms. Finally, the last host galaxy (top right panel - Galaxy 235) appears to be a clumpy proto-galaxy that has no distinct morphology as of yet and it is characterised by many high-density knots of gas. This clumpy case is the most common host for the infant GC candidates in Table \ref{table:formationproperties}. 
In the following Sections we will examine the impact that these different types of host galaxy environment might have on the formation and early evolution of the infant GC candidates.

\subsection{Disk-like galaxies}
\label{sec:rotation}

The {\it LL} and {\it ZDM} (see Table \ref{table:formationproperties}) candidates are located in host galaxies that show a clear disk structure and, for the latter, also spiral arms 
(see Figure \ref{fig:hostsystems}, bottom two panels). 
For the two  {\it ZDM} candidates (GCs 15 and 16), there is a difference in their formation times of about $\sim 17$ Myr, but they form close to each other and have similar masses and Jeans lengths. 
The candidates appear to form in one of the spiral arms of the host galaxy.
This is  the natural place for their formation given that  
spiral arms are regions rich in molecular gas clouds and are known to host high-density gas, as well as clustered star formation \citep[e.g., see][]{kim02,kim06,mo10}. Spiral arms themselves can trigger gravitational instabilities in the gas clouds \citep[e.g., see][]{elmegreen09}.


To further explore the possible role of the angular momentum of the host galaxy in defining the environment of the different classes of infant GC candidates, 
we evaluate $\lambda$, i.e. the global spin parameter of the host galaxies, computed  at the formation times of their GC candidates. We begin by following the method of \cite{lagos17} to calculate the radial $\lambda_R$ profile (see their Equations 1 - 4). In their work, such profiles only take into account the baryonic components; here, we wish to apply this method to both the baryons and the dark matter particles. To calculate the radius of each shell, we assume a Lagrangian-based approach, such that each shell contains the same number of particles ($N=4000$).
Once the radial profile is computed, the global $\lambda$ is found by summing the $\lambda_R$ value in each shell and then dividing the total by the number of shells. 

\begin{figure}
\centering
\includegraphics[width=\columnwidth]{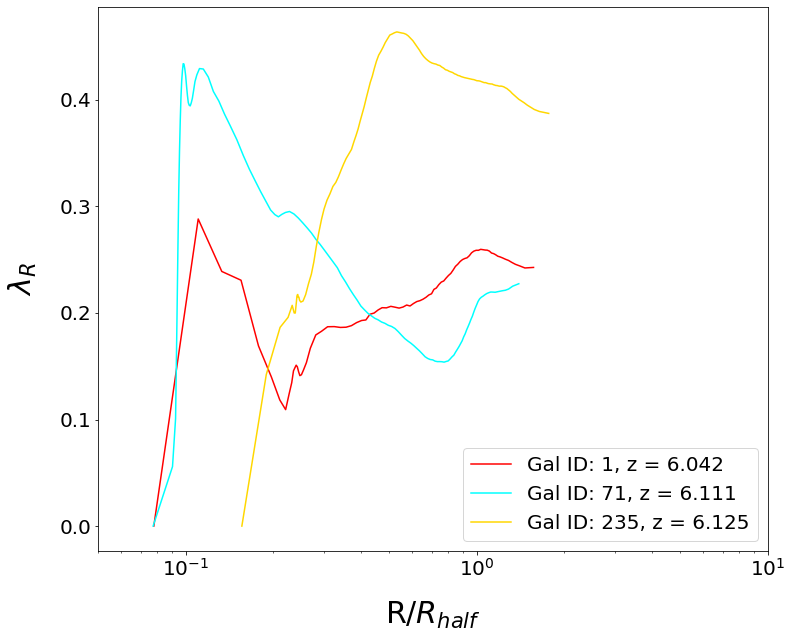}
\caption{Radial profile of the dimensionless spin parameter for three representative host galaxies of the main groups of GC candidates where the x-axis is normalised to the half-mass radius of the host galaxy. The host for {\it LL} is displayed in red, {\it ZDM} in blue, and {\it STD} in yellow (see Table \ref{table:formationproperties}). The redshift at which the profile is calculated is indicated in the legend.}
\label{fig:spinprofile}
\end{figure}

The global $\lambda$ value for both the galaxies in the proto-merging system (Galaxy 1) and the `standard' proto-galaxy (Galaxy 235) decreases with redshift, whereas the rotating spiral (Galaxy 71, bottom panels, Figure \ref{fig:hostsystems}) has a constant value for $\lambda$ across the formation redshifts of its infant GC candidates. This indicates that the angular momentum in this system is being conserved. 
However, the value of the total $\lambda$ of Galaxy 71 is not the highest out of the galaxies analysed here. The reason for this is clear when inspecting the $\lambda_R$ profiles (see Figure \ref{fig:spinprofile}). While baryons dominate the inner parts of the galaxies analysed here, the decision to include dark matter particles in the calculation has resulted in a flattening of the profile out to large radii for Galaxy 235, thus inflating the global value of $\lambda$. If we focus instead on the scales $< 1$ kpc  (which would pertain to the stellar disk for the spiral galaxy), it is clear that Galaxy 71 has a much higher value of angular momentum. 

The GC candidates with the largest range in angular momentum (class {\it LL}, see Table \ref{table:formationproperties}) happen to be in the same host galaxy (Galaxy 1).This galaxy also hosts one of the {\it SO} candidates, making it the galaxy with the most GC candidates of our study. This host also has a defined disk-like structure and, when moving out to larger scales, we find that it is about to engage in a galaxy-galaxy merger. We investigate now whether this future galaxy-galaxy merger already has an impact on the formation of infant GC candidates within this host galaxy. A number of  previous studies have indeed looked into the impact that interacting and merging galaxies could have on the formation of GCs \citep[e.g., see][]{kennicutt88,ashman92,whitmore95,whitmore99,bekki02,whitmore10,miah15, kim18}. 

\begin{figure}
\centering
\includegraphics[width=\columnwidth]{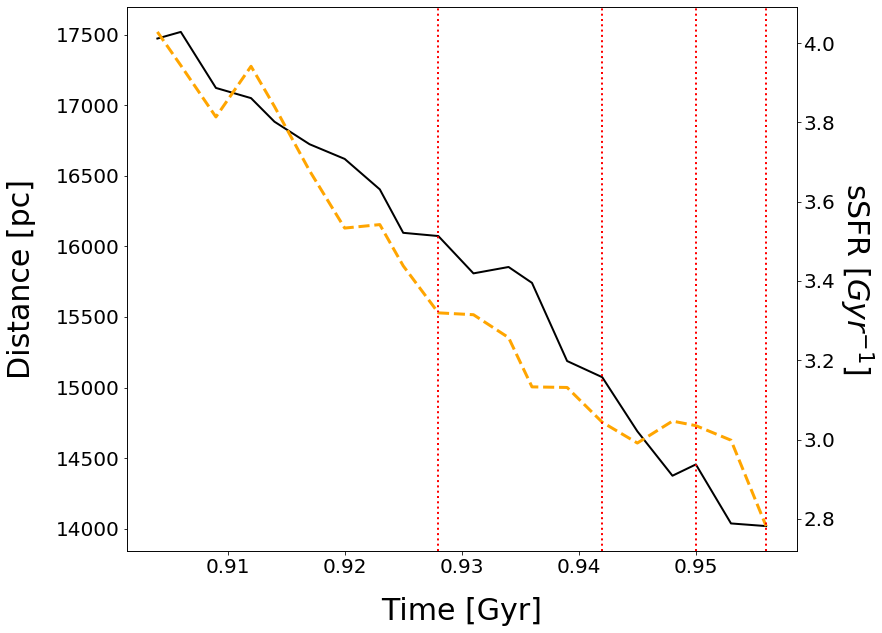}
\caption{Evolution of the sSFR of Galaxy 1 (orange, dashed) compared with its distance from Galaxy 0 (black, solid). The vertical lines indicate times where infant GC candidates belonging to Galaxy 1 `form'.} 
\label{fig:merginggalaxy}
\end{figure}

Although at a very early stage of the merger, the proximity between these two galaxies may result in a higher number of candidates for host Galaxy 1. To investigate this further, we examine the specific star formation rate (sSFR) across time for Galaxy 1 (orange dashed line in Figure \ref{fig:merginggalaxy}), as well as the evolution of the distance between the two galaxies (black solid line in Figure \ref{fig:merginggalaxy}). 
If the onset of the merger is having an impact on the star formation in Galaxy 1, we would expect to see spikes in the sSFR as the two galaxies approach each other \citep{springel05,dimatteo08}. 
We do not see evidence for the existence of such a correlation in Figure \ref{fig:merginggalaxy}, which suggests that the merger is still in its very early stages. There is no substantial orbital evolution and the two galaxies have not yet gone through their first pericenter passage, therefore, it would be premature to see the typical starburst event associated with the advanced stages of a merger. 

However, infant GC candidates start to form with higher frequency as the two galaxies approach each other, suggesting that instabilities are already triggered in the disc of Galaxy 1.
As it can be seen in Figure \ref{fig:hostsystems}, there is a bridge of gas connecting these two galaxies. 
An investigation at higher redshifts of the galaxies taking part in this event shows that the bridge begins to form before the formation of any of the candidates in Galaxy 1 (by more than one redshift). In future work, we will look into the effect that this assembling bridge may have on the existing candidates in this system, as well as the bridge's capacity to host further GC candidates. 

In summary, during the early stages of a major merger, before the onset of any starburst, we already find that the frequency of GC formation increases as the merging galaxies approach each other.


\subsection{Clumpy irregular proto-galaxies}
\label{sec:hostmass}
In the FiBY simulations, we find that most of the galaxies hosting GC candidates are not yet well-developed spiral galaxies, but rather ``standard'' high-redshift clumpy proto-galaxies with no distinct disk or bulge; they do consist of high-density clumps of gas within a structure of irregular morphology.
As we move to galaxies with fewer and fewer GC candidates, the hosts become less massive and more diffuse. When compared to the two other hosts previously discussed in Section \ref{sec:rotation}, these galaxies also appear to have lower stellar and gas masses, which helps explaining their dominance, due to the steep power-law slope of the stellar mass function. 

\begin{figure}
\centering    
\includegraphics[width=\columnwidth]{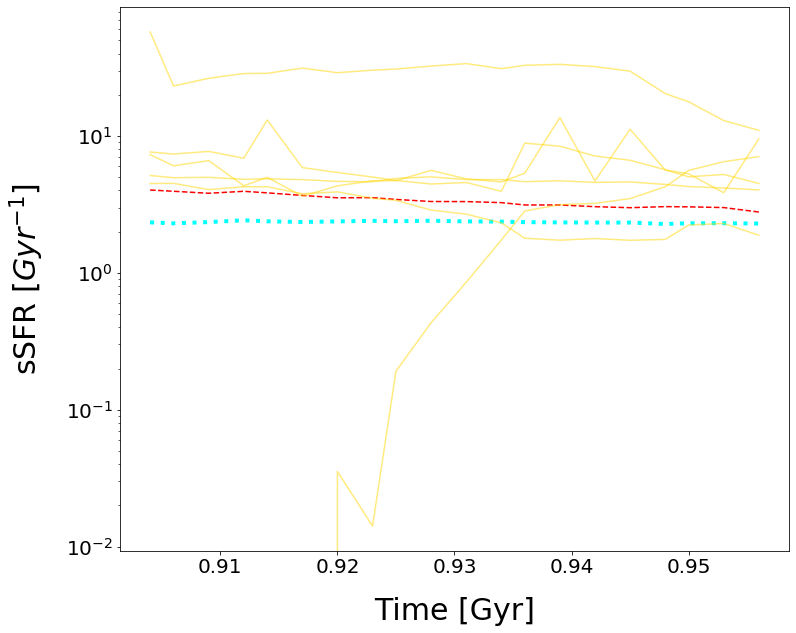}
\caption{The evolution of sSFR for all galaxies hosting candidates in the FiBY simulation. The clumpy high-redshift proto-galaxies (orange solid lines) in general have higher sSFR than the rotating spiral hosts (blue dotted line) or the disk galaxy involved in a merger (red dashed line). }
\label{fig:hostssfr}
\end{figure}

We analyse these high-redshift hosts in a similar fashion to the previous two cases (see Table 1 for the resulting properties). 
These hosts do not appear to have exceptionally large values  of the spin parameter, nor do they have any signs of previous or future mergers. We study the sSFR over time for all hosts considered in this work and we compare this quantity to the star formation histories of the infant GC candidates, as well as their ``formation times''. 
When analysing the sSFR of the host galaxies we recover our results from P1, where we established a relationship between the sSFR of the host galaxy and the mass of the most massive globular in a system (see Figure 9 of P1). 
In general, these clumpy host galaxies have higher sSFR (orange solid lines, Figure \ref{fig:hostssfr}) compared to  the rotating spiral hosts or the disk galaxy involved in a merger, by at least an order of magnitude. Such a difference in sSFR is reflected in the masses of the GC candidates belonging to these systems (see Table \ref{table:formationproperties}). For instance, candidates 15 and 16, which are located in the rotating spiral galaxy, have lower masses, on average, than the GC candidates in the clumpy hosts. 

In summary, we find that the sSFRs of our ``standard'' clumpy proto-galaxies are, in general, larger and more varied than those we find for the more disk-like hosts. These clumpy high-redshift galaxies, in turn, produce more massive infant GC candidates, consistent with our relation from P1, although they are typically lower in mass compared to the other host galaxies. 
This suggests that the morphology of the host is a good proxy for the mode of star formation as indicated by the sSFR, and, in turn, this mode affects the mass of individual infant GC candidates at formation.   


\begin{figure}
\centering
\includegraphics[width=\columnwidth]{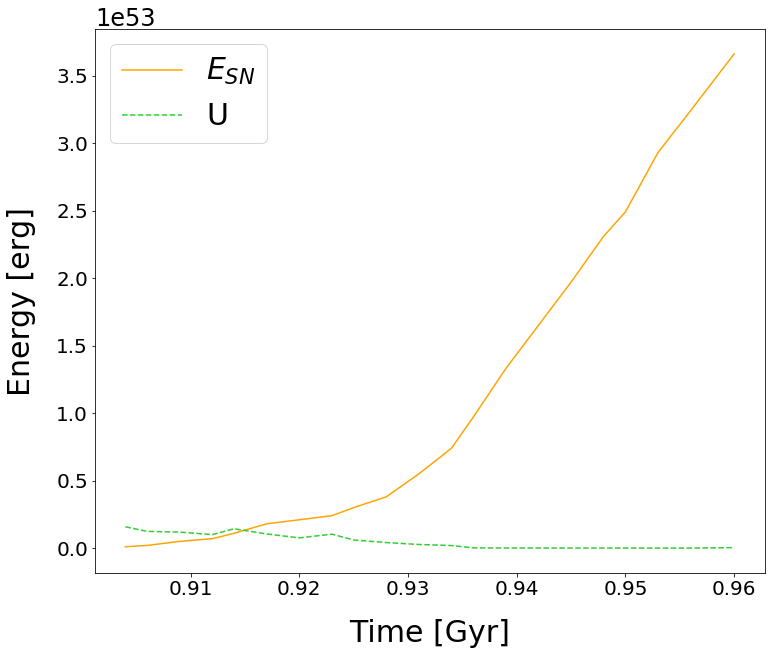}
\caption{Energy evolution of the particles within the half-mass radius of one of the {\it SO} candidates (GC 10). The green dashed line shows the potential energy of the gas particles in the half-mass radius and the orange line shows the cumulative SN energy.}
\label{fig:snenergy}
\end{figure}

\section{Feedback Effects and Subsequent Evolution}
\label{sec:staronly}

As previously anticipated, we identify two GC candidates, labelled as {\it SO}, that, at $z=6$, are composed exclusively of stars. They are, physically, the most similar to the GCs we observe in the local Universe. 
One of the most interesting features of the {\it SO} candidates evolution in phase space is the distinct lack of gas for most of the snapshots studied. At earlier redshifts, the gas appears to be gravitationally bound, but, suddenly, the energy of the gas increases substantially, and, eventually, all the gas is blown out of the half-mass radius. 
In our simulations, the source of this feedback energy are SN explosions. The prescription for SN feedback in the FiBY simulations (see Section \ref{sec:sims}) means that thermal energy is imparted to the surrounding gas, which, in turn, will increase the outward velocity of the gas in the vicinity of multiple SN. To evaluate whether the energy injected by SN could be enough to disrupt the gas component in these GC candidates and drive a wind, we compare the cumulative SN energy with the potential energy of the gas in the system over time. This is shown in Figure \ref{fig:snenergy}, where it appears that the cumulative energy of the SN at earlier times is smaller than the potential energy of the infant GC. As the number of SN increase with time, the energy deposited into the surrounding gas slowly increases. Eventually, it increases in an almost exponential fashion; by this point, the gas particles in the GC candidate have been blown out. The cross-over point past which the potential energy becomes significantly lower than the SN contribution coincides with the redshift where the gas particles in phase space rapidly increase in energy and become unbound from the infant GC. This shows that SN feedback is the main physical mechanism behind the gas depletion in this candidate \citep[see also][]{davis14} and not gas consumption by star formation. 

\begin{figure}
\centering
\includegraphics[width=\columnwidth]{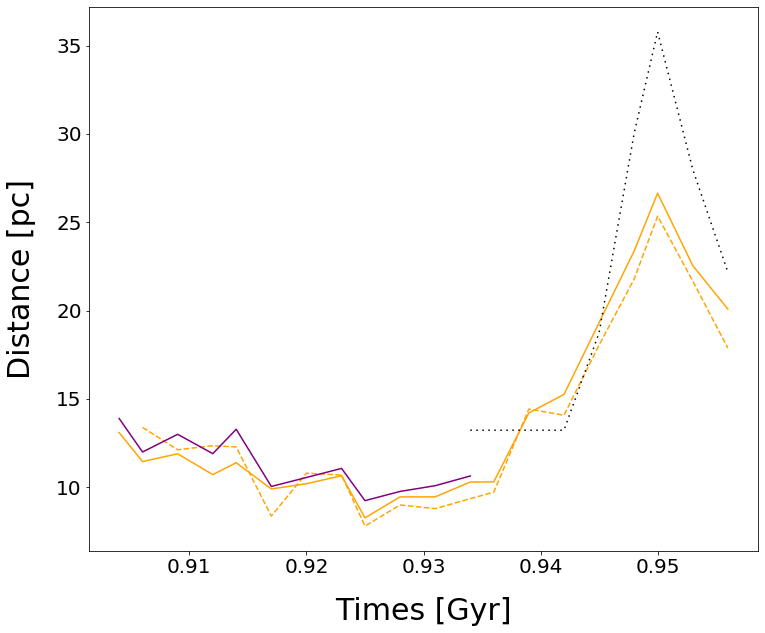}
\caption{Evolution of the average interparticle distance for the gas (purple) and star (orange) particles for one of the {\it SO} candidates (GC 10). All stars are shown with solid lines whilst stars that were present in the previous snapshot are shown with dashed lines to highlight the impact of newly formed stars between snapshots. The black dotted line shows the prediction due to adiabatic expansion given in Equation \ref{eqn:adiabatic}.}
\label{fig:adiabaticexpan}
\end{figure}

The removal of such a large quantity of mass in a short amount of time has an impact on the remaining star particles in the GC candidate. To investigate this further, in Figure \ref{fig:adiabaticexpan} we compute the interparticle distance for the gas (purple solid) and star (orange lines) particles. The gas and star interparticle distances are very similar at earlier times, which is consistent with the gas condensing and forming stars. The evolution of the gas abruptly ends at the time when the gas is expelled from the half-mass radius. After this moment, the interparticle distance between the stars increases; the removal of mass violently changes the potential of the infant GC candidate, thus causing the GC to expand.
To quantify this interpretation, we use the prescription of \cite{blumenthal86}, which has later been expressed by \cite{arraki14} as
\begin{equation} \label{eqn:adiabatic}
r_{\rm f} = \frac{r_{\rm i}}{\alpha(r_{\rm i})},
\end{equation}
where $r_{\rm i}$ is the radius of the object at the final snapshot before all the gas is expelled,  $\alpha(r_{\rm i})$ is the fraction of mass within this radius at a later time (once the gas has been removed) and $r_{\rm f}$ is the radius we expect at this time. In our calculations, $r_{\rm i}$ is constant,  but $\alpha(r_{\rm i})$ changes with redshift. By examining the theoretical prediction illustrated in Figure \ref{fig:adiabaticexpan} (dotted black line), the behaviour of the star particles in this GC candidate at later redshifts appears to be approximately consistent with an adiabatic expansion. While the average inter-particle distance shown in Figure \ref{fig:adiabaticexpan} is of the same order as the smoothing length and thus affected by smoothing of the potential on small scales, we show it to highlight that the expansion of the system in our simulation is due to feedback driven outflows. This conclusion should not be affected by the use of smoothed potentials as further supported by looking at Figure \ref{fig:snenergy} in which we show the potential energy under the assumption of point masses. Even in this case the energy from supernova feedback exceeds the potential energy by a factor of few.

In summary, in the {\it SO} candidates we find that the stellar feedback due to SN is strong enough to blow out the gas component. Once the gas is removed, there is a period of time during which the star particles resettle themselves into the new potential, and this appears to happen in an adiabatic fashion. We suggest that the gas-rich GC candidates that have been identified in other hosts will follow a similar evolution beyond $z=6$. The most likely reason why GC candidates 10 and 35 have already reached this more advanced dynamical stage is that they start forming stars earlier and, therefore, the associated stellar feedback is in effect sooner. 

\section{Discussion}
\label{sec:discuss}

\subsection{Implications of the infant GC\lowercase{s}-host connection}
We have presented a connection between the morphology of the host galaxy and the internal properties of the infant GC candidates during their early evolution. Host galaxies with little global angular momentum, but very clumpy morphology, appear to host infant GC candidates that have less internal rotation compared to those formed in more developed spiral galaxies (see Section 5). These results suggest an avenue to generate two populations of GCs with different internal properties, as a consequence of their host environments at formation. It is interesting to note that galaxies with a high number of clumps and velocity dispersion \citep[e.g.][]{2009ApJ...706.1364F,2009ApJ...700L..21K} are expected to  be common at higher redshifts due to the high accretion rates of gas; only beyond $z \sim 2$ a population of disk-like galaxies starts appearing. 

If GC formation is indeed a natural by-product of ongoing star formation \citep[e.g., see][]{kruijssen15}, the FiBY simulations would suggest that the GC candidates that are progressively formed at later redshifts will be characterised by a higher internal angular momentum content, as their host will become more regular and less clumpy.
The sSFR of the clumpy hosts appear to be, on average, higher than those of the more disk-like galaxies; this is consistent with the fact that clumpy hosts also display higher accretion rates per unit mass and it suggests a higher specific GC formation rate. The evidence for the latter is only tentative, as it is based on the relatively limited statistics of the present study. 

\subsection{Impact of stellar feedback and gas expulsion}
The impact of stellar feedback on high-density gas regions (e.g. for molecular clouds in merging galaxies) has been found to be less efficient at clearing the gas, thus high-density gas clouds could have multiple episodes of star formation \citep{rahner17,guillard18}.
To test this idea, we examined the evolution of the gas morphology of the galaxy hosting the two SO candidates, before and after the expulsion of gas from the candidates.

For candidate 10, we see that, at earlier times, the object appears as a dense knot of gas within a tightly furled spiral arm. As time progresses, the galaxy rotates and the candidate moves with the spiral arm. In the few snapshots before the gas expulsion from this object, the knot of gas appears to become gradually less dense. This behaviour is also consistent with what we observe for the behaviour of the gas in the Lindblad diagram for this object. After $z = 6.111$ (the redshift at which the gas is expelled from this candidate), the spiral arm appears to ``unfurl'' and a cavity of low-density gas grows around the location of the GC candidate. 
This suggests that such a cavity might be a result of stellar feedback from the infant GC candidate. It is known that stellar feedback in molecular cloud structures can determine the expulsion of gas - in some cases removing it completely - which, in turn, can halt star formation \citep{murray10,krumholz14,hopkins14}; gas expulsion can also affect the density structure of the surrounding environment \citep{grisdale17,semenov17}. 


When examining the evolution of the environment surrounding candidate 35, we find that the infant GC appears to maintain a position near the center of the clumpy proto-galaxy. 
Prior to the redshift at which the gas is expelled from candidate 35, we note only minor variations in the morphology of the gas of the host. After that moment, we note an expansion of the high-density core and also the appearance of two low-density cavities on either side of the location of the GC candidate. 
To summarise, after the moment in which the gas is expelled from the {\it SO} candidates, there appears to be also a change in the density of the gas within the host galaxy. 

\subsection{Future observational studies}

A significant finding from this work is the confirmation of the relation between the specific star formation rate of the host galaxy and the mass of the most massive GC in the system. As presented in \citet{phipps20}, this relation appears to hold true at low and high redshift, for the simulated and observational data sets we studied. The fact that the relation persists in the high-frequency FiBY simulation offers evidence that this result is impervious to the stochastic modelling of star formation between different numerical realisations. Such a relation can be utilised in current and forthcoming observations of high-redshift galaxies to predict the masses of the largest infant GCs expected to form at a given redshift; therefore, it can offer a guide to identify particularly promising host galaxies to target in future observational campaigns. 

\subsection{Numerical caveats}
The evidence from the FiBY simulation used in this work is that feedback due to SN is efficient at removing gas from the young stellar clusters and hence halting star formation. However, in FiBY the only type of stellar feedback prescribed is that due to SN. We have not included the effects of other types of stellar feedback that could be happening early in the infant GC candidates lives. The inclusion of other types of feedback could  have an effect on our results as it has been shown that some forms of stellar feedback can have impact on the gas cloud / stellar cluster before the first SNe occur. In the work of \cite{grudic22}, they study a very high-resolution simulation of an individual giant molecular cloud from its initial collapse until its dispersal. The simulation includes protostellar jets, radiation, winds and SN. They find that during star formation, the jets are the dominant form of feedback but they cannot disrupt the gas cloud on their own. Once winds and radiation from sufficiently massive stars comes into play the cloud is successfully disrupted and star formation is halted. This occurs on a timescale of $\sim 8$ Myr but the first SN does not occur until $\sim 8.3$ Myr into the simulation displaying that in their work if you include early forms of stellar feedback, the feedback from SN is redundant when it comes to halting star formation. Similar results were presented in \cite{verliat22}, who found that ionising radiation was needed to disperse the gas cloud but protostellar jets were important in the regulation of star formation. These simulations, however, did not include the effects of stellar winds or SN. Based on the recent works in the literature, it is likely that including these early stellar feedback mechanisms would have an effect on the way the infant GC candidates in FiBY evolve in the early stages, however it must be noted that the resolutions used in these recent works \citep{grudic22,verliat22} are much finer and they only focus on a singular isolated gas cloud hence they are able to resolve the formation of individual stars. In the future, we would consider incorporating other forms of feedback into the FiBY simulations and comparing the results to the ones presented here. 

The radii of the infant GC candidates studied in this work are similar in size to the gravitational softening length of the FiBY simulations. Due to this, we decided to check whether the point-mass based method of calculating the gravitational energy used in this work was still feasible. To do this, we recalculated the gravitational potential energies for the candidates using Equations 70 and 71 of \cite{spring01} and compared the values retrieved with that of the point-mass based method. We found very minimal changes in the values for gravitational potential energy, with regards to Figure \ref{fig:redsixlindblad}, we observed a slight upward shift in the energies of the particles when using the Equations from \cite{spring01} but this shift is small enough that it does not change the interpretation of our results. Any differences were primarily due to plotting normalised values for energy, therefore we conclude that the point-mass method used in this work is valid even with the small sizes of the candidates. 

\section{Conclusions}
\label{sec:conclusion}
We have built upon the previous results from \cite{phipps20} and utilised a high-frequency snapshot ($\Delta t \sim$ 2 Myr)
simulation from the First Billion Years project to explore the possible formation channels of a group of low-mass stellar systems which are likely ancestors of the low-redshift globular clusters - we call these stellar systems ``infant GC candidates''. The main results of this study can be summarised as follows:

\begin{itemize}

\item We harnessed the behaviour of the infant GC candidates in the energy and angular momentum plane (``Lindblad diagram'') as a technique to examine the formation of small structures in a changing galactic environment (Figure \ref{fig:redsixlindblad}). This analysis of the phase space evolution over time allowed us to identify the redshift at which these systems first became gravitationally bound (``formation redshift''), as well as to show evidence of radial infall of gas onto the object at early times (Figure \ref{fig:lindevol}). 

\item Before this ``formation redshift'', the gas in the infant GC candidates collapses on a dynamical timescale. Further analysis of the infall of the star particles revealed that the ratio of the Jeans length of gas to the interparticle distance between stars is of order unity, which suggests that the size of infant GCs during the early stages is set by the Jeans length of the gas. 

\item The phase space behaviour of the infant GC candidates allowed us to identify some characteristic groupings of objects (Table \ref{table:formationproperties}), which may indicate the existence of multiple formation channels. Such a phase space classification is also reflected in the formation environment: GC candidates which belong to the same class are located in host galaxies of similar morphology (Table \ref{table:hostproperties} and Figure \ref{fig:hostsystems}) 


\item The majority of the infant GC candidates (i.e., those with lower angular momentum) are located in ``standard'' clumpy high-redshift proto-galaxies which have no distinct morphology yet. 
Furthermore, we found that the relation between the sSFR of the host galaxy and the mass of the most massive candidate in a system \citep[established in][]{phipps20} persisted here, supporting the argument that these candidates are a natural product of star formation in high-redshift galaxies.  

\item Finally, we examined the potential future evolution for the general population of infant GC candidates through the analysis of two GC candidates that contains only stars by $z=6$. Such candidates most closely resemble the GCs we observe in the local Universe. We found that, at earlier times, the feedback energy from SN expells the gas from the half-mass radius of the systems and, subsequently, the remaining star particles relax into the new potential via an approximately adiabatic expansion.
\end{itemize}

This latter point will be investigated further in the future, by means of tailored numerical simulations which extend beyond $z=6$. Specifically, we wish to assess whether the evolutionary path discussed in Section \ref{sec:staronly} is common for the majority of the GC candidates. Most importantly, it will be essential to first evaluate if these infant GC candidates will survive until $z=0$ and how closely they will resemble to present-day GCs, as observed in the local Universe. 

\section*{Acknowledgements}

FP is supported by an STFC Studentship (Ref.~2145045) and a Saltire Early Career Fellowship from the Royal Society of Edinburgh; she is grateful to the Instituto de Astrof{\'i}sica de Canarias for the kind hospitality in Winter 2022. ALV acknowledges support from a UKRI Future Leaders Fellowship (MR/S018859/1). CDV has been supported by the Spanish Ministry of Science and Innovation
(MICIU/FEDER) through research grant RYC-2015-18078 and PGC2018-094975-C22. For the purpose of open access, the authors have applied a Creative Commons Attribution (CC BY) licence to any Author Accepted Manuscript version arising from this submission. \newline

\section*{Data Availability}
The data used in this work is available upon reasonable request.





\bibliographystyle{mnras}
\bibliography{paper2bibfile.bib}

\begin{thebibliography}{}
\makeatletter
\relax
\def\mn@urlcharsother{\let\do\@makeother \do\$\do\&\do\#\do\^\do\_\do\%\do\~}
\def\mn@doi{\begingroup\mn@urlcharsother \@ifnextchar [ {\mn@doi@}
  {\mn@doi@[]}}
\def\mn@doi@[#1]#2{\def\@tempa{#1}\ifx\@tempa\@empty \href
  {http://dx.doi.org/#2} {doi:#2}\else \href {http://dx.doi.org/#2} {#1}\fi
  \endgroup}
\def\mn@eprint#1#2{\mn@eprint@#1:#2::\@nil}
\def\mn@eprint@arXiv#1{\href {http://arxiv.org/abs/#1} {{\tt arXiv:#1}}}
\def\mn@eprint@dblp#1{\href {http://dblp.uni-trier.de/rec/bibtex/#1.xml}
  {dblp:#1}}
\def\mn@eprint@#1:#2:#3:#4\@nil{\def\@tempa {#1}\def\@tempb {#2}\def\@tempc
  {#3}\ifx \@tempc \@empty \let \@tempc \@tempb \let \@tempb \@tempa \fi \ifx
  \@tempb \@empty \def\@tempb {arXiv}\fi \@ifundefined
  {mn@eprint@\@tempb}{\@tempb:\@tempc}{\expandafter \expandafter \csname
  mn@eprint@\@tempb\endcsname \expandafter{\@tempc}}}

\bibitem[\protect\citeauthoryear{Abel, Anninos, Zhang  \& Norman}{Abel
  et~al.}{1997}]{abel97}
Abel T.,  Anninos P.,  Zhang Y.,   Norman M.~L.,  1997, New Astronomy, 2, 181

\bibitem[\protect\citeauthoryear{Agarwal, Dalla~Vecchia, Johnson, Khochfar  \&
  Paardekooper}{Agarwal et~al.}{2014}]{agarwal14}
Agarwal B.,  Dalla~Vecchia C.,  Johnson J.~L.,  Khochfar S.,   Paardekooper
  J.~P.,  2014, MNRAS, 443, 648

\bibitem[\protect\citeauthoryear{Arraki, Klypin, More  \&
  Trujillo-Gomez}{Arraki et~al.}{2014}]{arraki14}
Arraki K.~S.,  Klypin A.,  More S.,   Trujillo-Gomez S.,  2014, MNRAS, 438,
  1466

\bibitem[\protect\citeauthoryear{Ashman \& Zepf}{Ashman \&
  Zepf}{1992}]{ashman92}
Ashman K.~M.,  Zepf S.~E.,  1992, ApJ, 384, 50

\bibitem[\protect\citeauthoryear{Bastian \& Lardo}{Bastian \&
  Lardo}{2018}]{bastian18}
Bastian N.,  Lardo C.,  2018, ARA\&A, 56, 83

\bibitem[\protect\citeauthoryear{Bastian, Lamers, de Mink, Longmore, Goodwin
  \& Gieles}{Bastian et~al.}{2013}]{bastian13}
Bastian N.,  Lamers H. J. G. L.~M.,  de Mink S.~E.,  Longmore S.~N.,  Goodwin
  S.~P.,   Gieles M.,  2013, MNRAS, 436, 2398

\bibitem[\protect\citeauthoryear{Bate, Bonnell  \& Bromm}{Bate
  et~al.}{2003}]{bate03}
Bate M.~R.,  Bonnell I.~A.,   Bromm V.,  2003, MNRAS, 339, 577

\bibitem[\protect\citeauthoryear{Beasley, Baugh, Forbes, Sharples  \&
  Frenk}{Beasley et~al.}{2002}]{beasley02}
Beasley M.~A.,  Baugh C.~M.,  Forbes D.~A.,  Sharples R.~M.,   Frenk C.~S.,
  2002, MNRAS, 333, 383

\bibitem[\protect\citeauthoryear{Behroozi, Wechsler  \& Wu}{Behroozi
  et~al.}{2013}]{behroozi13}
Behroozi P.~S.,  Wechsler R.~H.,   Wu H.~Y.,  2013, ApJ, 762

\bibitem[\protect\citeauthoryear{Bekki}{Bekki}{2017}]{bekki17}
Bekki K.,  2017, MNRAS, 467, 1857

\bibitem[\protect\citeauthoryear{Bekki, Forbes, Beasley  \& Couch}{Bekki
  et~al.}{2002}]{bekki02}
Bekki K.,  Forbes D.~A.,  Beasley M.~A.,   Couch W.~J.,  2002, MNRAS, 335, 1176

\bibitem[\protect\citeauthoryear{Blumenthal, Faber, Flores  \&
  Primack}{Blumenthal et~al.}{1986}]{blumenthal86}
Blumenthal G.~R.,  Faber S.~M.,  Flores R.,   Primack J.~R.,  1986, ApJ, 301,
  27

\bibitem[\protect\citeauthoryear{Bouwens, Illingworth, Oesch, Maseda, Ribeiro,
  Stefanon  \& Lam}{Bouwens et~al.}{2017}]{bouwens17}
Bouwens R.~J.,  Illingworth G.~D.,  Oesch P.~A.,  Maseda M.,  Ribeiro B.,
  Stefanon M.,   Lam D.,  2017, eprint

\bibitem[\protect\citeauthoryear{{Bouwens}, {Illingworth}, {van Dokkum},
  {Ribeiro}, {Oesch}  \& {Stefanon}}{{Bouwens} et~al.}{2021}]{bouwens21}
{Bouwens} R.~J.,  {Illingworth} G.~D.,  {van Dokkum} P.~G.,  {Ribeiro} B.,
  {Oesch} P.~A.,   {Stefanon} M.,  2021, \mn@doi [\aj]
  {10.3847/1538-3881/abfda6}, \href
  {https://ui.adsabs.harvard.edu/abs/2021AJ....162..255B} {162, 255}

\bibitem[\protect\citeauthoryear{{Bouwens}, {Illingworth}, {van Dokkum},
  {Oesch}, {Stefanon}  \& {Ribeiro}}{{Bouwens} et~al.}{2022}]{bouwens22}
{Bouwens} R.~J.,  {Illingworth} G.~D.,  {van Dokkum} P.~G.,  {Oesch} P.~A.,
  {Stefanon} M.,   {Ribeiro} B.,  2022, \mn@doi [\apj]
  {10.3847/1538-4357/ac4791}, \href
  {https://ui.adsabs.harvard.edu/abs/2022ApJ...927...81B} {927, 81}

\bibitem[\protect\citeauthoryear{Bullock, Dekel, Kolaff, Kravtsov, Klypin  \&
  Porciani}{Bullock et~al.}{2001}]{bullock01}
Bullock J.~S.,  Dekel A.,  Kolaff T.~S.,  Kravtsov A.~V.,  Klypin A.~A.,
  Porciani C.,  2001, ApJ, 555, 240

\bibitem[\protect\citeauthoryear{Caffau et~al.,}{Caffau
  et~al.}{2011}]{caffau11}
Caffau E.,  et~al., 2011, Nature, 477, 67

\bibitem[\protect\citeauthoryear{Chabrier}{Chabrier}{2003}]{chabrier03}
Chabrier G.,  2003, PASP, 115, 763

\bibitem[\protect\citeauthoryear{C{\^o}t{\'e}, Marzke  \& West}{C{\^o}t{\'e}
  et~al.}{1998}]{cote98}
C{\^o}t{\'e} P.,  Marzke R.~O.,   West M.~J.,  1998, ApJ, 501, 554

\bibitem[\protect\citeauthoryear{Creasey, Sales, Peng  \& Sameie}{Creasey
  et~al.}{2019}]{creasey19}
Creasey P.,  Sales L.~V.,  Peng E.~W.,   Sameie O.,  2019, MNRAS, 482, 219

\bibitem[\protect\citeauthoryear{Cullen, McLure, Khochfar, Dunlop  \&
  Dalla~Vecchia}{Cullen et~al.}{2017}]{cullen17}
Cullen F.,  McLure R.~J.,  Khochfar S.,  Dunlop J.~S.,   Dalla~Vecchia C.,
  2017, MNRAS, 470, 3006

\bibitem[\protect\citeauthoryear{D'Ercole, Vesperini, D'Antona, McMillan  \&
  Recchi}{D'Ercole et~al.}{2008}]{dercole08}
D'Ercole A.,  Vesperini E.,  D'Antona F.,  McMillan S. L.~W.,   Recchi S.,
  2008, MNRAS, 391, 825

\bibitem[\protect\citeauthoryear{D'Ercole, D'Antona  \& Vesperini}{D'Ercole
  et~al.}{2016}]{dercole16}
D'Ercole A.,  D'Antona F.,   Vesperini E.,  2016, MNRAS, 461, 4088

\bibitem[\protect\citeauthoryear{Dalla~Vecchia \& Schaye}{Dalla~Vecchia \&
  Schaye}{2012}]{dallavecchia12}
Dalla~Vecchia C.,  Schaye J.,  2012, MNRAS, 426, 140

\bibitem[\protect\citeauthoryear{Davis, Khochfar  \& Dalla~Vecchia}{Davis
  et~al.}{2014}]{davis14}
Davis A.~J.,  Khochfar S.,   Dalla~Vecchia C.,  2014, MNRAS, 443, 985

\bibitem[\protect\citeauthoryear{Decressin, Meynet, Charbonnel, Prantzos  \&
  Ekstrom}{Decressin et~al.}{2007a}]{decressin07a}
Decressin T.,  Meynet G.,  Charbonnel C.,  Prantzos N.,   Ekstrom S.,  2007a,
  A\&A, 464, 1029

\bibitem[\protect\citeauthoryear{Decressin, Charbonnel  \& Meynet}{Decressin
  et~al.}{2007b}]{decressin07b}
Decressin T.,  Charbonnel C.,   Meynet G.,  2007b, A\&A, 475, 859

\bibitem[\protect\citeauthoryear{Di~Matteo, Bournaud, Martig, Combes, Melchior
  \& Semelin}{Di~Matteo et~al.}{2008}]{dimatteo08}
Di~Matteo P.,  Bournaud F.,  Martig M.,  Combes F.,  Melchior A.~L.,   Semelin
  B.,  2008, A\&A, 492, 31

\bibitem[\protect\citeauthoryear{Dobbs, Krumholz, Ballesteros-Paredes, Bolatto,
  Fukui, Heyer, Low  \& Ostriker}{Dobbs et~al.}{2014}]{dobbs14}
Dobbs C.~L.,  Krumholz M.~R.,  Ballesteros-Paredes J.,  Bolatto A.~D.,  Fukui
  Y.,  Heyer M.,  Low M. M.~M.,   Ostriker E.~C.,  2014, Protostars and
  Planets, pp 3--26

\bibitem[\protect\citeauthoryear{Dolag, Borgani, Murante  \& Springel}{Dolag
  et~al.}{2009}]{dolag09}
Dolag K.,  Borgani S.,  Murante G.,   Springel V.,  2009, MNRAS, 399, 497

\bibitem[\protect\citeauthoryear{Elmegreen}{Elmegreen}{1979}]{elmegreen79}
Elmegreen B.~G.,  1979, ApJ, 231, 372

\bibitem[\protect\citeauthoryear{Elmegreen}{Elmegreen}{1995}]{elmegreen95}
Elmegreen B.~G.,  1995, Molecular Couds and Star Formation, p.~149

\bibitem[\protect\citeauthoryear{Elmegreen}{Elmegreen}{2009}]{elmegreen09}
Elmegreen B.~G.,  2009, The Galaxy Disk in Cosmological Context, Proceedings of
  the International Astronomical Union, IAU Symposium, 254, 289

\bibitem[\protect\citeauthoryear{Elmegreen}{Elmegreen}{2017}]{elmegreen17a}
Elmegreen B.~G.,  2017, ApJ, 836, 80

\bibitem[\protect\citeauthoryear{Elmegreen \& Elmegreen}{Elmegreen \&
  Elmegreen}{2017}]{elmegreen17b}
Elmegreen D.~M.,  Elmegreen B.~G.,  2017, ApJL, 851, L44

\bibitem[\protect\citeauthoryear{Ferland, Korista, Verner, Ferguson, Kingdon
  \& Verner}{Ferland et~al.}{1998}]{ferland98}
Ferland G.~J.,  Korista K.~T.,  Verner D.~A.,  Ferguson J.~W.,  Kingdon J.~B.,
   Verner E.~M.,  1998, PASP, 110, 761

\bibitem[\protect\citeauthoryear{Field \& Salslaw}{Field \&
  Salslaw}{1965}]{field65b}
Field G.~B.,  Salslaw W.~C.,  1965, ApJ, 142, 568

\bibitem[\protect\citeauthoryear{Forbes \& Bridges}{Forbes \&
  Bridges}{2010}]{forbes10}
Forbes D.~A.,  Bridges T.,  2010, MNRAS, 404, 1203

\bibitem[\protect\citeauthoryear{Forbes, Brodie  \& Grillmair}{Forbes
  et~al.}{1997}]{forbes97}
Forbes D.~A.,  Brodie J.~P.,   Grillmair C.~J.,  1997, AJ, 113, 1652

\bibitem[\protect\citeauthoryear{{F{\"o}rster Schreiber} et~al.,}{{F{\"o}rster
  Schreiber} et~al.}{2009}]{2009ApJ...706.1364F}
{F{\"o}rster Schreiber} N.~M.,  et~al., 2009, \mn@doi [\apj]
  {10.1088/0004-637X/706/2/1364}, \href
  {https://ui.adsabs.harvard.edu/abs/2009ApJ...706.1364F} {706, 1364}

\bibitem[\protect\citeauthoryear{Frebel, Johnson  \& Bromm}{Frebel
  et~al.}{2007}]{frebel07}
Frebel A.,  Johnson J.~L.,   Bromm V.,  2007, MNRAS, 380, L40

\bibitem[\protect\citeauthoryear{Galli \& Palla}{Galli \&
  Palla}{1998}]{galli98}
Galli D.,  Palla F.,  1998, A\&A, 335, 403

\bibitem[\protect\citeauthoryear{Grisdale, Agertz, Romeo, Renaud  \&
  Read}{Grisdale et~al.}{2017}]{grisdale17}
Grisdale K.,  Agertz O.,  Romeo A.~B.,  Renaud F.,   Read J.~I.,  2017, MNRAS,
  466, 1093

\bibitem[\protect\citeauthoryear{Grudi\'c, Guszejnov, Offner, Rosen, Raju,
  Faucher-Giguere  \& Hopkins}{Grudi\'c et~al.}{2022}]{grudic22}
Grudi\'c M.~Y.,  Guszejnov D.,  Offner S. S.~R.,  Rosen A.~L.,  Raju A.~N.,
  Faucher-Giguere C.~A.,   Hopkins P.~F.,  2022, MNRAS, 512, 216

\bibitem[\protect\citeauthoryear{Guillard, Emsellen  \& Renaud}{Guillard
  et~al.}{2018}]{guillard18}
Guillard N.,  Emsellen E.,   Renaud F.,  2018, MNRAS, 477, 5001

\bibitem[\protect\citeauthoryear{Halbesma, Grand, Gomez, Marinacci, Pakmor,
  Trick, Busch  \& White}{Halbesma et~al.}{2020}]{halbesma20}
Halbesma T L.~R.,  Grand R. J.~J.,  Gomez F.~A.,  Marinacci F.,  Pakmor R.,
  Trick W.~H.,  Busch P.,   White S. D.~M.,  2020, MNRAS, 496, 638

\bibitem[\protect\citeauthoryear{Heggie \& Hut}{Heggie \& Hut}{2003}]{heggie03}
Heggie D.,  Hut P.,  2003, The Gravitational Million-Body Problem: A
  Multidisciplinary Approach To Star Cluster Dynamics.
Cambridge University Press

\bibitem[\protect\citeauthoryear{Hopkins, Kere\v{s}, O\~norbe,
  Faucher-Gigu\`ere, Quataert, Murray  \& Bullock}{Hopkins
  et~al.}{2014}]{hopkins14}
Hopkins P.~F.,  Kere\v{s} D.,  O\~norbe J.,  Faucher-Gigu\`ere C.~A.,  Quataert
  E.,  Murray N.,   Bullock J.~S.,  2014, MNRAS, 445, 581

\bibitem[\protect\citeauthoryear{Howard, Pudritz  \& Harris}{Howard
  et~al.}{2018}]{howard18}
Howard C.~S.,  Pudritz R.~E.,   Harris W.~E.,  2018, NatAs, 2, 725

\bibitem[\protect\citeauthoryear{Johnson, Dalla~Vecchia  \& Khochfar}{Johnson
  et~al.}{2013}]{johnson13}
Johnson J.~L.,  Dalla~Vecchia C.,   Khochfar S.,  2013, MNRAS, 428, 1857

\bibitem[\protect\citeauthoryear{Katz \& Ricotti}{Katz \&
  Ricotti}{2013}]{katz13}
Katz H.,  Ricotti M.,  2013, MNRAS, 432, 3250

\bibitem[\protect\citeauthoryear{Kawamata, Ishigaki, Shimasaku, Oguri, Ouchi
  \& Tarigawa}{Kawamata et~al.}{2018}]{kawamata18}
Kawamata R.,  Ishigaki M.,  Shimasaku K.,  Oguri M.,  Ouchi M.,   Tarigawa S.,
  2018, ApJ, 855, 4

\bibitem[\protect\citeauthoryear{Keller, Kruijssen, Pfeffer, Reina-Campos,
  Bastian, Trujillo-Gomez, Hughes  \& Crain}{Keller et~al.}{2020}]{keller20}
Keller B.~W.,  Kruijssen J. M.~D.,  Pfeffer J.,  Reina-Campos M.,  Bastian N.,
  Trujillo-Gomez S.,  Hughes M.~E.,   Crain R.~A.,  2020, MNRAS, 495, 4248

\bibitem[\protect\citeauthoryear{Kennicutt}{Kennicutt}{1998}]{kennicutt98}
Kennicutt R. C.~J.,  1998, AR\&A, 36, 189

\bibitem[\protect\citeauthoryear{Kennicutt \& Chu}{Kennicutt \&
  Chu}{1988}]{kennicutt88}
Kennicutt R C.~J.,  Chu Y.-H.,  1988, AJ, 95, 720

\bibitem[\protect\citeauthoryear{{Khochfar} \& {Burkert}}{{Khochfar} \&
  {Burkert}}{2001}]{2001ApJ...561..517K}
{Khochfar} S.,  {Burkert} A.,  2001, \mn@doi [\apj] {10.1086/323382}, \href
  {https://ui.adsabs.harvard.edu/abs/2001ApJ...561..517K} {561, 517}

\bibitem[\protect\citeauthoryear{{Khochfar} \& {Silk}}{{Khochfar} \&
  {Silk}}{2009}]{2009ApJ...700L..21K}
{Khochfar} S.,  {Silk} J.,  2009, \mn@doi [\apjl]
  {10.1088/0004-637X/700/1/L21}, \href
  {https://ui.adsabs.harvard.edu/abs/2009ApJ...700L..21K} {700, L21}

\bibitem[\protect\citeauthoryear{Kikuchihara et~al.,}{Kikuchihara
  et~al.}{2020}]{kikuchihara20}
Kikuchihara S.,  et~al., 2020, ApJ, 893, 60

\bibitem[\protect\citeauthoryear{Kim \& Ostriker}{Kim \&
  Ostriker}{2002}]{kim02}
Kim W.~T.,  Ostriker E.~C.,  2002, ApJ, 570, 132

\bibitem[\protect\citeauthoryear{Kim, Kim  \& Ostriker}{Kim
  et~al.}{2006}]{kim06}
Kim C.~G.,  Kim W.~T.,   Ostriker E.~C.,  2006, ApJ, 649, L13

\bibitem[\protect\citeauthoryear{Kim et~al.,}{Kim et~al.}{2018}]{kim18}
Kim J.~H.,  et~al., 2018, MNRAS, 474, 4232

\bibitem[\protect\citeauthoryear{Knebe et~al.,}{Knebe et~al.}{2011}]{knebe11}
Knebe A.,  et~al., 2011, MNRAS, 415, 2293

\bibitem[\protect\citeauthoryear{Knollmann \& Knebe}{Knollmann \&
  Knebe}{2009}]{knollmann09}
Knollmann S.~R.,  Knebe A.,  2009, ApJs, 182, 608

\bibitem[\protect\citeauthoryear{Kravtsov \& Gnedin}{Kravtsov \&
  Gnedin}{2005}]{kravtsov05}
Kravtsov A.~V.,  Gnedin O.~Y.,  2005, ApJ, 623, 650

\bibitem[\protect\citeauthoryear{Kruijssen}{Kruijssen}{2015}]{kruijssen15}
Kruijssen J. M.~D.,  2015, MNRAS, 454, 1658

\bibitem[\protect\citeauthoryear{Kruijssen, Pfeffer, Crain  \&
  Bastian}{Kruijssen et~al.}{2019}]{kruijssen19}
Kruijssen J. M.~D.,  Pfeffer J.~L.,  Crain R.~A.,   Bastian N.,  2019, MNRAS,
  486, 3134

\bibitem[\protect\citeauthoryear{Krumholz}{Krumholz}{2014}]{krumholz14}
Krumholz M.~R.,  2014, PhR, 539, 49

\bibitem[\protect\citeauthoryear{Kwan}{Kwan}{1979}]{kwan79}
Kwan J.,  1979, ApJ, 229, 567

\bibitem[\protect\citeauthoryear{Lagos, Theuns, Stevens, Cortese, Padilla,
  Davis, Contreras  \& Croton}{Lagos et~al.}{2017}]{lagos17}
Lagos C.~P.,  Theuns T.,  Stevens A. R.~H.,  Cortese L.,  Padilla N.~D.,  Davis
  T.~A.,  Contreras S.,   Croton D.,  2017, MNRAS, 464, 3850

\bibitem[\protect\citeauthoryear{Lardo, Bellazzini, Pancino, Carretta,
  Bragaglia  \& Dalessandro}{Lardo et~al.}{2011}]{lardo11}
Lardo C.,  Bellazzini M.,  Pancino E.,  Carretta E.,  Bragaglia A.,
  Dalessandro E.,  2011, A\&A, 525, 11

\bibitem[\protect\citeauthoryear{Larsen, Brodie, Huchra, Forbes  \&
  Grillmair}{Larsen et~al.}{2001}]{larsen01}
Larsen S.~S.,  Brodie J.~P.,  Huchra J.~P.,  Forbes D.~A.,   Grillmair C.~J.,
  2001, AJ, 121, 2974

\bibitem[\protect\citeauthoryear{{Larson}}{{Larson}}{1969}]{1969MNRAS.145..271L}
{Larson} R.~B.,  1969, \mn@doi [\mnras] {10.1093/mnras/145.3.271}, \href
  {https://ui.adsabs.harvard.edu/abs/1969MNRAS.145..271L} {145, 271}

\bibitem[\protect\citeauthoryear{Leaman, VandenBerg  \& Mendel}{Leaman
  et~al.}{2013}]{leaman13}
Leaman R.,  VandenBerg D.~A.,   Mendel J.~T.,  2013, MNRAS, 436, 122

\bibitem[\protect\citeauthoryear{Li \& Gnedin}{Li \& Gnedin}{2019}]{li19}
Li H.,  Gnedin O.~Y.,  2019, MNRAS, 486, 4030

\bibitem[\protect\citeauthoryear{Ma et~al.,}{Ma et~al.}{2020}]{ma2020}
Ma X.,  et~al., 2020, MNRAS, 493, 4315

\bibitem[\protect\citeauthoryear{Maio, Khochfar, Johnson  \& Ciardi}{Maio
  et~al.}{2011}]{maio11}
Maio U.,  Khochfar S.,  Johnson J.~L.,   Ciardi B.,  2011, MNRAS, 414, 1145

\bibitem[\protect\citeauthoryear{McKee \& Ostriker}{McKee \&
  Ostriker}{2007}]{mckee07}
McKee C.~F.,  Ostriker E.~C.,  2007, ARA\&A, 45, 565

\bibitem[\protect\citeauthoryear{{Me{\v{s}}tri{\'c}}
  et~al.,}{{Me{\v{s}}tri{\'c}} et~al.}{2022}]{mevstric22}
{Me{\v{s}}tri{\'c}} U.,  et~al., 2022, \mn@doi [arXiv e-prints]
  {10.48550/arXiv.2202.09377}, \href
  {https://ui.adsabs.harvard.edu/abs/2022arXiv220209377M} {p. arXiv:2202.09377}

\bibitem[\protect\citeauthoryear{Miah, Sharples  \& Cho}{Miah
  et~al.}{2015}]{miah15}
Miah J.~A.,  Sharples R.~M.,   Cho J.,  2015, MNRAS, 447, 3639

\bibitem[\protect\citeauthoryear{Milone et~al.,}{Milone
  et~al.}{2017}]{milone17}
Milone A.~P.,  et~al., 2017, MNRAS, 464, 3636

\bibitem[\protect\citeauthoryear{Mo, van~den Bosch  \& White}{Mo
  et~al.}{2010}]{mo10}
Mo H.,  van~den Bosch F.~C.,   White S.,  2010, Galaxy Formation and Evolution.
Cambridge University Press, Cambridge, UK

\bibitem[\protect\citeauthoryear{Murray, Quataert  \& Thompson}{Murray
  et~al.}{2010}]{murray10}
Murray N.,  Quataert E.,   Thompson T.~A.,  2010, ApJ, 709, 191

\bibitem[\protect\citeauthoryear{Nakasato, Mori  \& Nomoto}{Nakasato
  et~al.}{2000}]{nakasato00}
Nakasato N.,  Mori M.,   Nomoto K.,  2000, ApJ, 535, 776

\bibitem[\protect\citeauthoryear{Neistein, Li, Khochfar, Weinmann, Shankar  \&
  Boylan-Kolchin}{Neistein et~al.}{2011}]{neistein11}
Neistein E.,  Li C.,  Khochfar S.,  Weinmann S.~M.,  Shankar F.,
  Boylan-Kolchin M.,  2011, MNRAS, 416, 1486

\bibitem[\protect\citeauthoryear{Omukai}{Omukai}{2000}]{omukai00}
Omukai K.,  2000, ApJ, 534, 809

\bibitem[\protect\citeauthoryear{{Paardekooper}, {Khochfar}  \&
  {Dalla}}{{Paardekooper} et~al.}{2013}]{2013MNRAS.429L..94P}
{Paardekooper} J.~P.,  {Khochfar} S.,   {Dalla} C.~V.,  2013, \mn@doi [\mnras]
  {10.1093/mnrasl/sls032}, \href
  {https://ui.adsabs.harvard.edu/abs/2013MNRAS.429L..94P} {429, L94}

\bibitem[\protect\citeauthoryear{Paardekooper, Khochfar  \&
  Dalla~Vecchia}{Paardekooper et~al.}{2015}]{paardekooper15}
Paardekooper J.~P.,  Khochfar S.,   Dalla~Vecchia C.,  2015, MNRAS, 451, 2544

\bibitem[\protect\citeauthoryear{Parker}{Parker}{1966}]{parker66}
Parker E.~N.,  1966, ApJ, 145, 811

\bibitem[\protect\citeauthoryear{Peebles}{Peebles}{1969}]{peebles69}
Peebles P. J.~E.,  1969, ApJ, 155

\bibitem[\protect\citeauthoryear{Peebles \& Dicke}{Peebles \&
  Dicke}{1968}]{peebles68}
Peebles P. J.~E.,  Dicke R.~H.,  1968, ApJ, 154, 891

\bibitem[\protect\citeauthoryear{Peng et~al.,}{Peng et~al.}{2006}]{peng06}
Peng E.~W.,  et~al., 2006, ApJ, 639, 95

\bibitem[\protect\citeauthoryear{{Penston}}{{Penston}}{1969}]{1969MNRAS.144..425P}
{Penston} M.~V.,  1969, \mn@doi [\mnras] {10.1093/mnras/144.4.425}, \href
  {https://ui.adsabs.harvard.edu/abs/1969MNRAS.144..425P} {144, 425}

\bibitem[\protect\citeauthoryear{Pfeffer, Kruijssen, Crain  \& Bastian}{Pfeffer
  et~al.}{2018}]{pfeffer18}
Pfeffer J.,  Kruijssen J. M.~D.,  Crain R.~A.,   Bastian N.,  2018, MNRAS, 475,
  4309

\bibitem[\protect\citeauthoryear{Phipps, Khochfar, Varri  \&
  Dalla~Vecchia}{Phipps et~al.}{2020}]{phipps20}
Phipps F.,  Khochfar S.,  Varri A.~L.,   Dalla~Vecchia C.,  2020, A\&A, 641,
  A132

\bibitem[\protect\citeauthoryear{Piotto et~al.,}{Piotto
  et~al.}{2015}]{piotto15}
Piotto G.,  et~al., 2015, AJ, 149, 28

\bibitem[\protect\citeauthoryear{Plummer}{Plummer}{1911}]{plummer11}
Plummer H.~C.,  1911, MNRAS, 71, 460

\bibitem[\protect\citeauthoryear{Prantzos \& Charbonnel}{Prantzos \&
  Charbonnel}{2006}]{prantzos06}
Prantzos N.,  Charbonnel C.,  2006, A\&A, 458, 135

\bibitem[\protect\citeauthoryear{Prieto \& Gnedin}{Prieto \&
  Gnedin}{2008}]{prieto08}
Prieto J.~L.,  Gnedin O.~Y.,  2008, ApJ, 689, 919

\bibitem[\protect\citeauthoryear{Rahner, Pelligrini, Glover  \& Klessen}{Rahner
  et~al.}{2017}]{rahner17}
Rahner D.,  Pelligrini E.~W.,  Glover S. C.~O.,   Klessen R.~S.,  2017, MNRAS,
  470, 4453

\bibitem[\protect\citeauthoryear{Recio-Blanco}{Recio-Blanco}{2018}]{recioblanco18}
Recio-Blanco A.,  2018, A\&A, 620, A194

\bibitem[\protect\citeauthoryear{Renaud, Agertz  \& Gieles}{Renaud
  et~al.}{2017}]{renaud17}
Renaud F.,  Agertz O.,   Gieles M.,  2017, MNRAS, 465, 3622

\bibitem[\protect\citeauthoryear{Ricotti, Parry  \& Gnedin}{Ricotti
  et~al.}{2016}]{ricotti16}
Ricotti M.,  Parry O.~H.,   Gnedin N.~Y.,  2016, ApJ, 831, 204

\bibitem[\protect\citeauthoryear{Salpeter}{Salpeter}{1955}]{salpeter55}
Salpeter E.~E.,  1955, ApJ, 121, 161

\bibitem[\protect\citeauthoryear{Schaye \& Dalla~Vecchia}{Schaye \&
  Dalla~Vecchia}{2008}]{schaye08}
Schaye J.,  Dalla~Vecchia C.,  2008, MNRAS, 383, 1210

\bibitem[\protect\citeauthoryear{Schaye et~al.,}{Schaye
  et~al.}{2010}]{schaye10}
Schaye J.,  et~al., 2010, MNRAS, 402, 1136

\bibitem[\protect\citeauthoryear{Schmidt}{Schmidt}{1959}]{schmidt59}
Schmidt M.,  1959, ApJ, 129, 243

\bibitem[\protect\citeauthoryear{Schweizer}{Schweizer}{1987}]{schweizer87}
Schweizer F.,  1987, Nearly Normal Galaxies. From the Planck Time to the
  Present. The Eighth Santa Cruz Summer Workshop in Astronomy and Astrophysics,
  p.~18

\bibitem[\protect\citeauthoryear{Semenov, Kravtsov  \& Gnedin}{Semenov
  et~al.}{2017}]{semenov17}
Semenov V.~A.,  Kravtsov A.~V.,   Gnedin N.~Y.,  2017, ApJ, 845, 133

\bibitem[\protect\citeauthoryear{Springel}{Springel}{2005}]{spring05}
Springel V.,  2005, MNRAS, 364, 1105

\bibitem[\protect\citeauthoryear{Springel, Yoshida  \& White}{Springel
  et~al.}{2001}]{spring01}
Springel V.,  Yoshida N.,   White S. D.~M.,  2001, New Astronomy, 6, 79

\bibitem[\protect\citeauthoryear{Springel, Di~Matteo  \& Hernquist}{Springel
  et~al.}{2005}]{springel05}
Springel V.,  Di~Matteo T.,   Hernquist L.,  2005, MNRAS, 361, 776

\bibitem[\protect\citeauthoryear{Tonini}{Tonini}{2013}]{tonini13}
Tonini C.,  2013, ApJ, 762, 11

\bibitem[\protect\citeauthoryear{Tornatore, Borgani, Dolag  \&
  Matteucci}{Tornatore et~al.}{2007}]{tornatore07}
Tornatore L.,  Borgani S.,  Dolag K.,   Matteucci F.,  2007, MNRAS, 382, 1050

\bibitem[\protect\citeauthoryear{Vanzella et~al.,}{Vanzella
  et~al.}{2017}]{vanzella17}
Vanzella E.,  et~al., 2017, MNRAS, 467, 4304

\bibitem[\protect\citeauthoryear{Vanzella et~al.,}{Vanzella
  et~al.}{2019}]{vanzella19}
Vanzella E.,  et~al., 2019, MNRAS, 483, 3618

\bibitem[\protect\citeauthoryear{Vanzella et~al.,}{Vanzella
  et~al.}{2020}]{vanzella20}
Vanzella E.,  et~al., 2020, MNRAS, 491, 1093

\bibitem[\protect\citeauthoryear{Verliat, Hennebelle, Gonz\'alez, Lee  \&
  Geen}{Verliat et~al.}{2022}]{verliat22}
Verliat A.,  Hennebelle P.,  Gonz\'alez M.,  Lee Y.~N.,   Geen S.,  2022, A\&A,
  663, A6

\bibitem[\protect\citeauthoryear{Whitmore \& Schweizer}{Whitmore \&
  Schweizer}{1995}]{whitmore95}
Whitmore B.~C.,  Schweizer F.,  1995, AJ, 109, 960

\bibitem[\protect\citeauthoryear{Whitmore, Zhang, Leitherer, Fall, Schweizer
  \& Miller}{Whitmore et~al.}{1999}]{whitmore99}
Whitmore B.~C.,  Zhang Q.,  Leitherer C.,  Fall S.~M.,  Schweizer F.,   Miller
  B.~W.,  1999, AJ, 118, 1551

\bibitem[\protect\citeauthoryear{Whitmore et~al.,}{Whitmore
  et~al.}{2010}]{whitmore10}
Whitmore B.~C.,  et~al., 2010, AJ, 140, 75

\bibitem[\protect\citeauthoryear{Wiersma, Schaye  \& Smith}{Wiersma
  et~al.}{2009a}]{wiersma09}
Wiersma R. P.~C.,  Schaye J.,   Smith B.~D.,  2009a, MNRAS, 393, 99

\bibitem[\protect\citeauthoryear{Wiersma, Schaye, Theuns, Dalla~Vecchia  \&
  Tornatore}{Wiersma et~al.}{2009b}]{wiersma09b}
Wiersma R. P.~C.,  Schaye J.,  Theuns T.,  Dalla~Vecchia C.,   Tornatore L.,
  2009b, MNRAS, 399, 574

\bibitem[\protect\citeauthoryear{Yoshida, Omukai, Hernquist  \& Abel}{Yoshida
  et~al.}{2006}]{yoshida06}
Yoshida N.,  Omukai K.,  Hernquist L.,   Abel T.,  2006, ApJ, 652, 6

\makeatother
\end{thebibliography}
\bsp	
\label{lastpage}
\end{document}